\documentclass[10pt,conference,letterpaper]{IEEEtran}
\pdfoutput=1
\usepackage{amsmath,graphicx,amssymb,bm,pstcol}
\usepackage[flushleft]{threeparttable}
\usepackage{makecell}
\usepackage[linesnumbered,ruled,vlined]{algorithm2e}
\usepackage[caption=false,font=footnotesize]{subfig}
\usepackage{balance}

\usepackage{verbatim}
\usepackage[hyphens]{url}
\usepackage{multirow}
\usepackage[bookmarks=true,hidelinks]{hyperref}
\usepackage{hypcap}
\usepackage{xcolor}
\hypersetup{
	colorlinks,
	linkcolor={red!50!black},
	citecolor={blue!50!black},
	urlcolor={blue!80!black}
}

\newcommand{\numberthis}{\stepcounter{equation}\tag{\theequation}}

\newtheorem{theorem}{Theorem}

\newtheorem{lemma}{Lemma}

\newcommand{\figuresname}{Figs.}

\graphicspath{{figures/}}
\author{
	\IEEEauthorblockN{Jing Tang}
	\IEEEauthorblockA{Interdisciplinary Graduate School\\
		Nanyang Technological University\\
		Email: tang0311@ntu.edu.sg}
	\and
	\IEEEauthorblockN{Xueyan Tang}
	\IEEEauthorblockA{School of Comp. Sci. \& Eng.\\
		Nanyang Technological University\\
		Email: asxytang@ntu.edu.sg}
	\and
	\IEEEauthorblockN{Junsong Yuan}
	\IEEEauthorblockA{School of EEE\\
		Nanyang Technological University\\
		Email: jsyuan@ntu.edu.sg}
}

\title{Influence Maximization Meets Efficiency and Effectiveness: A Hop-Based Approach}
\begin{document}
\maketitle
\thispagestyle{plain}
\pagestyle{plain}
\begin{abstract}
Influence Maximization is an extensively-studied problem that targets at selecting a set of initial seed nodes in the Online Social Networks (OSNs) to spread the influence as widely as possible. However, it remains an open challenge to design fast and accurate algorithms to find solutions in large-scale OSNs. Prior Monte-Carlo-simulation-based methods are slow and not scalable, while other heuristic algorithms do not have any theoretical guarantee and they have been shown to produce poor solutions for quite some cases. In this paper, we propose hop-based algorithms that can easily scale to millions of nodes and billions of edges. Unlike previous heuristics, our proposed hop-based approaches can provide certain theoretical guarantees. Experimental evaluations with real OSN datasets demonstrate the efficiency and effectiveness of our algorithms.
\end{abstract}

\section{Introduction}\label{sec:introduction}
Information can be disseminated widely and rapidly through Online Social Networks (OSNs) with ``word-of-mouth'' effects. Viral marketing is such a typical application in which new products or activities are advertised by some influential users in the OSN to other users in a cascading manner \cite{Domingos_maxInfluence_2001}. A large amount of recent work \cite{Borgs_RIS_2014,Chen_MIA_2010,Chen_degreeDiscount_2009,Chen_LDAG_2010,Cheng_IMRank_2014,Cohen_SKIM_2014,Jung_IRIE_2012,Kempe_maxInfluence_2003,Leskovec_CELF_2007,Nguyen_BCT_2016,Nguyen_DSSA_2016,Ohsaka_prunedMC_2014,Tang_IMM_2015,Tang_reverse_2014,Zhou_UBLF_2013} has been focusing on \textit{influence maximization} in viral marketing, which targets at selecting a set of initial seed nodes in the OSN to spread the influence as widely as possible. The influence maximization problem was formulated by \cite{Kempe_maxInfluence_2003} with two basic diffusion models, namely the \textit{Independent Cascade} (IC) and \textit{Linear Threshold} (LT) models. Although finding the optimal seed set is NP-hard \cite{Kempe_maxInfluence_2003}, a simple greedy hill-climbing algorithm has a $(1-1/e)$-approximation guarantee due to the submodularity and monotone properties of the influence spread under these models \cite{Nemhauser_submodular_1978}. Follow-up studies have mostly concentrated on efficient implementation of the hill-climbing algorithm for large-scale OSNs. The key difference among various methods lies in how to estimate the influence spread of a seed set. 

Computing the exact influence spread on general graphs is \#P-hard for both the IC and LT models \cite{Chen_MIA_2010,Chen_LDAG_2010}. Thus, some Monte-Carlo-simulation-based methods \cite{Kempe_maxInfluence_2003,Leskovec_CELF_2007,Ohsaka_prunedMC_2014,Zhou_UBLF_2013} estimate the influence spread by reachability tests, while some \textit{reverse influence sampling} methods carry out seed selection \cite{Borgs_RIS_2014,Nguyen_BCT_2016,Nguyen_DSSA_2016,Tang_IMM_2015,Tang_reverse_2014} by leveraging the concept of reverse reachability. These sampling-based methods can provide theoretical guarantees up to $(1-1/e-\epsilon)$-approximation. However, the sampling-based methods can encounter the efficiency problem even for the dramatically improved versions \cite{Borgs_RIS_2014,Nguyen_BCT_2016,Nguyen_DSSA_2016,Ohsaka_prunedMC_2014,Tang_IMM_2015,Tang_reverse_2014} as they may consume a lot of time/memory to obtain/store one sample. Other heuristic methods \cite{Chen_MIA_2010,Chen_degreeDiscount_2009,Chen_LDAG_2010,Cheng_IMRank_2014,Jung_IRIE_2012} conduct rough estimation of the influence spread either by exploiting some related features (such as node degrees) or extracting subgraphs where the influence spread is easier to compute. However, these heuristics do not have any theoretical guarantee and they have been shown to suffer from the effectiveness problem.

To achieve both efficiency and effectiveness, in this paper, we propose a new hop-based approach for the influence maximization problem. Although it belongs to heuristics, unlike other heuristic methods, our method can provide certain theoretical guarantees. As shall be shown by the experimental results, our hop-based methods outperform the existing heuristics and perform as well as the sampling-based methods in terms of the influence spread produced. Meanwhile, our hop-based methods run much faster than the sampling-based methods. For a large OSN with billions of edges, only our hop-based methods and some ineffective heuristics can work with acceptable time and memory usage for various distributions of propagation probabilities. Our contributions are summarized as follows.
\begin{enumerate}
	\item We propose hop-based influence estimation algorithms for efficiently selecting seed nodes to maximize the influence spread.
	\item We develop an upper bounding approach on the influence generated by a seed node to further speed up seed selection.
	\item We carry out theoretical analysis for our hop-based algorithms and derive approximation guarantees.
	\item We conduct extensive experiments with several real OSN datasets. The results demonstrate the efficiency and effectiveness of our hop-based algorithms. 
\end{enumerate}

The rest of this paper is organized as follows. Section \ref{sec:problem} introduces the influence maximization problem and the greedy hill-climbing algorithm. Section \ref{sec:algorithms} elaborates our algorithm design and analyzes the theoretical guarantees. Section \ref{sec:evaluation} presents the experimental study. Section \ref{sec:relatedWork} reviews the related work. Finally, Section \ref{sec:conclusion} concludes the paper.

\section{Preliminaries}\label{sec:problem}

\subsection{Problem Definition}\label{subsec:preliminaries}

Let $\mathcal{G}=(\mathcal{V},\mathcal{E})$ be a directed graph modeling an OSN, where the nodes $\mathcal{V}$ represent users and the edges $\mathcal{E}$ represent the connections among users (e.g., followships on Twitter). For each directed edge $(u,v)\in \mathcal{E}$, we refer to $v$ as a \emph{neighbor} of $u$, and refer to $u$ as an \emph{inverse neighbor} of $v$.

To facilitate the exposition, we shall mainly focus on the Independent Cascade (IC) model --- a representative and most widely-studied diffusion model for influence propagation
\cite{Borgs_RIS_2014,Chen_MIA_2010,Chen_degreeDiscount_2009,Jung_IRIE_2012,Kempe_maxInfluence_2003,Leskovec_CELF_2007,Nguyen_BCT_2016,Nguyen_DSSA_2016,Ohsaka_prunedMC_2014,Tang_IMM_2015,Tang_reverse_2014,Zhou_UBLF_2013}. Extending our hop-based methods to other diffusion models shall be discussed later. In the IC model, a propagation probability $p_{u,v}$ is associated with each edge $(u,v)$, representing the probability for $v$ to be activated by $u$ through the edge. Let $\mathcal{N}_u$ denote the set of node $u$'s neighbors, i.e., $\mathcal{N}_u=\{v: v\in \mathcal{V}, (u,v)\in \mathcal{E}\}$. Given a set of seed nodes $\mathcal{S}$, the IC diffusion process proceeds as follows. Initially, the seed nodes $\mathcal{S}$ are activated, while all the other nodes are not activated. When a node $u$ first becomes activated, it attempts to further activate its neighbors who are not yet activated. For each such neighbor $v\in\mathcal{N}_u$, $v$ would become activated with probability $p_{u,v}$. This process repeats until no more node can be activated. The \textit{influence spread} of the seed set $\mathcal{S}$, denoted by $\sigma(\mathcal{S})$, is the expected number of nodes activated by the above process. The \textit{influence maximization} problem \cite{Kempe_maxInfluence_2003} is to find a set $\mathcal{S}$ of $k$ nodes to maximize $\sigma(\mathcal{S})$, where $k$ is a given parameter. Formally,
\begin{equation}\label{eq:problem}
\max_{|\mathcal{S}|=k}{\sigma(\mathcal{S})}.
\end{equation}

\subsection{Greedy Heuristic}\label{subsec:greedy}

The influence function $\sigma(\cdot)$ has been proved to be submodular and monotone under the IC model \cite{Kempe_maxInfluence_2003}. Thus, a simple greedy hill-climbing algorithm that provides $(1-1/e)$-approximation \cite{Nemhauser_submodular_1978} was proposed for influence maximization as described in Algorithm \ref{alg:greedy}. It starts with an empty seed set $\mathcal{S}=\emptyset$. In each iteration, the greedy heuristic chooses a new seed $u$ from the non-seed nodes $\mathcal{V}\setminus \mathcal{S}$ with largest marginal influence gain $\sigma(\mathcal{S}\cup\{u\})-\sigma(\mathcal{S})$ and adds $u$ to $\mathcal{S}$. The algorithm stops after selecting $k$ seeds. 

\begin{algorithm}[!h]
	\capstart
	\label{alg:greedy}
	\caption{$\bm{Greedy}(\mathcal{G}, \sigma)$}
	initialize $\mathcal{S}\leftarrow \emptyset$\;
	\While{the size of $\mathcal{S}$ is smaller than $k$}{
		find $u\leftarrow\arg\max_{v\in \mathcal{V}\setminus \mathcal{S}}\left\{\sigma(\mathcal{S}\cup\{v\})-\sigma(\mathcal{S})\right\}$\;
		$\mathcal{S}\leftarrow \mathcal{S}\cup \{u\}$\;
	}
	\Return $\mathcal{S}$\;
\end{algorithm}

A CELF technique \cite{Leskovec_CELF_2007} can be used to enhance the efficiency of the greedy algorithm due to the submodularity of influence spread. Specifically, it may not be necessary to evaluate the marginal influence gain for every node of $\mathcal{V}\setminus\mathcal{S}$ in each iteration. Due to the submodularity, the marginal gains can only decrease over iterations. Thus, the marginal gains obtained in the previous iterations can be used as upper bounds for a new iteration. In the new iteration, the nodes can be evaluated in decreasing order of these upper bounds. Once the largest marginal gain evaluated is greater than the upper bound of the next node to evaluate, the evaluation can stop as none of the remaining nodes would be able to produce larger marginal gain.

\section{Hop-Based Approaches}\label{sec:algorithms}

\subsection{Hop-Based Influence Estimation under IC Model}\label{subsec:hop}

\begin{figure}[!t]
	\capstart
	\centering
	\subfloat[High degree nodes]{\includegraphics[width=0.5\linewidth]{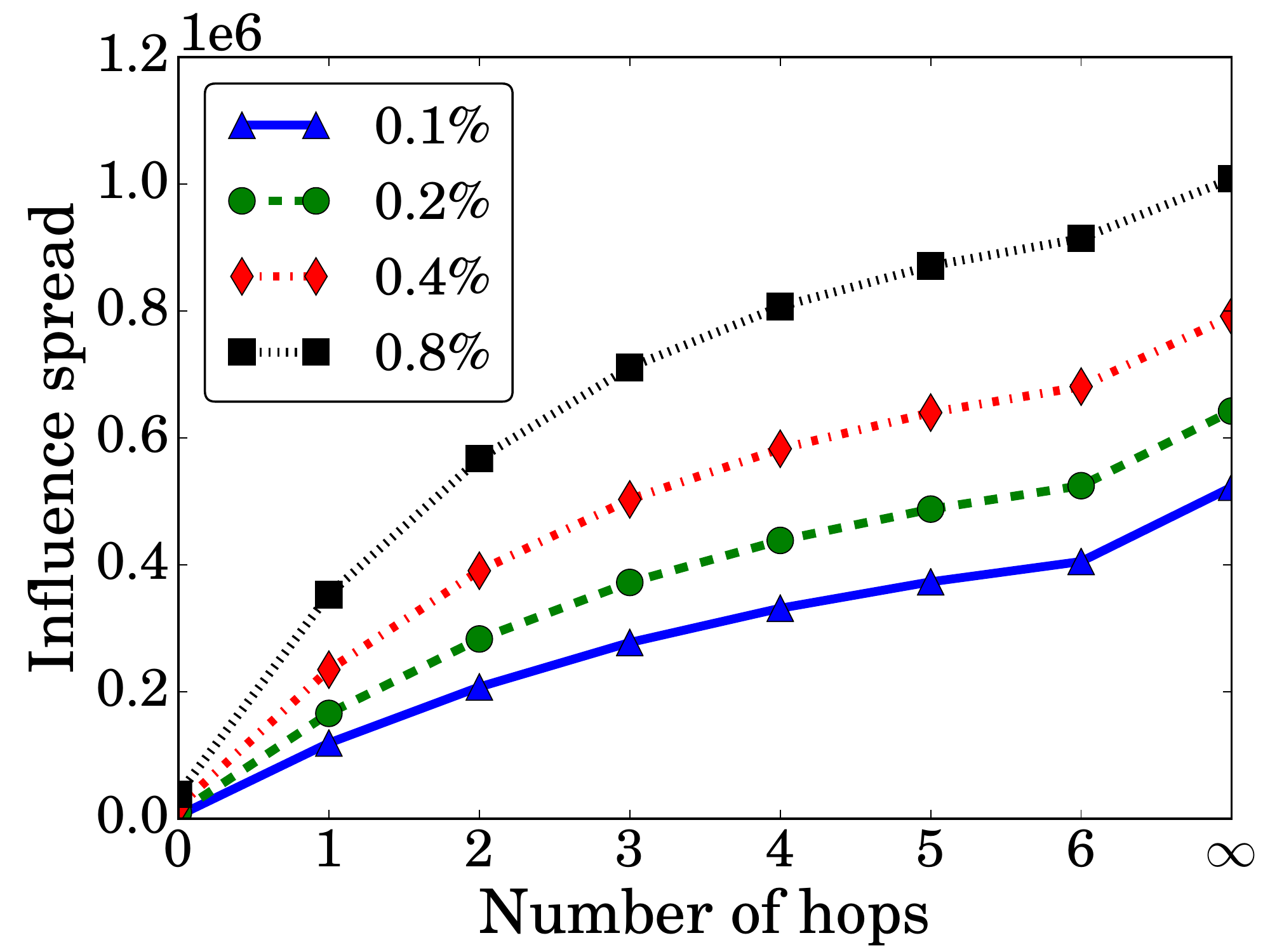}\label{subfig:liveJournal_highdegree}}\hfill
	\subfloat[Top influential nodes]{\includegraphics[width=0.5\linewidth]{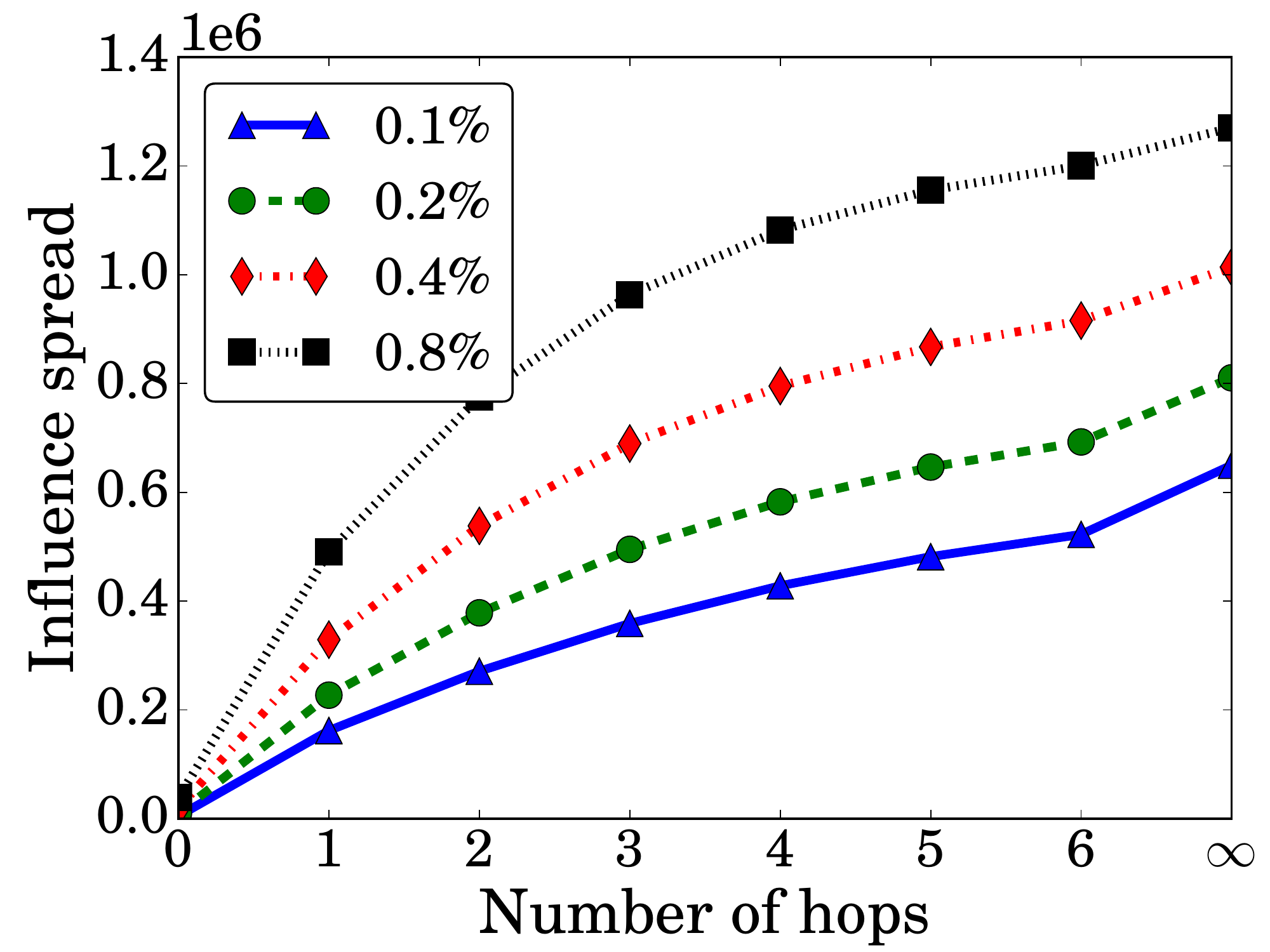}\label{subfig:liveJournal_top}}
	\caption{Influence spread for different hops of propagation on LiveJournal.}\label{fig:hopSampling}
	\vspace{-0.1in}
\end{figure}

We start by studying some empirical results of influence spread tested on a real OSN graph -- LiveJournal (5M nodes and 69M edges) \cite{snapnets}. \figurename~\ref{fig:hopSampling} plots the influence spread within different numbers of hops of propagation (the data points of $\infty$ hops represent the actual influence spread without any hop limit, and the seed set size is represented as a percentage of the entire node set in the OSN). We test two typical seed sets for influence maximization: selecting the nodes with highest degrees and selecting the top influential nodes using a greedy hill-climbing heuristic \cite{Kempe_maxInfluence_2003} as described in Section~\ref{subsec:greedy}. The propagation probability on each edge $(u,v)$ in the OSN is set to the reciprocal of $v$'s in-degree as widely adopted in previous studies \cite{Chen_degreeDiscount_2009,Kempe_maxInfluence_2003,Nguyen_DSSA_2016,Tang_IMM_2015}. As seen from \figurename~\ref{fig:hopSampling}, the increase in the influence spread for considering each additional hop of propagation generally decreases with increasing number of hops. The majority of influence spread is produced within the first few hops of propagation. Similar trends have also been observed by several measurement-driven studies on real OSNs \cite{Cha_Flickr_2009,Goel_structure_2012,Leskovec_viralMarketing_2007}. For example, Goel \textit{et al.} \cite{Goel_structure_2012} showed that less than $10\%$ of the cascades in the diffusion are more than $2$ hops away from the seed. These observations motivate us to design hop-based algorithms to efficiently capture the major influence propagation, especially for the first two hops of propagation.

A hop-based algorithm focuses on the influence propagation up to a given number of $h$ hops starting from the initial seed set. For $h=1$ and $h=2$, we can efficiently calculate the \textit{exact} influence spread within $h$ hops of propagation and maintain it incrementally when the seed set expands. Corresponding to the denotation of a node $v$'s neighbors as $\mathcal{N}_v$, let $\mathcal{I}_v$ denote $v$'s inverse neighbors, i.e., $\mathcal{I}_v=\{w:w\in\mathcal{V}, (w,v)\in\mathcal{E}\}$. Let $\pi_h^{\mathcal{S}}(v)$ denote the probability for a node $v$ to be activated within $h$ hops of propagation from a seed set $\mathcal{S}$, and let $\sigma_h(\mathcal{S})$ denote the influence spread produced within $h$ hops of propagation from $\mathcal{S}$.

\textbf{One Hop of Propagation:} We first model one hop of propagation. Obviously, for all the seed nodes $v\in\mathcal{S}$, $\pi_1^{\mathcal{S}}(v)=1$. With one hop of propagation, for all the non-seed nodes $v\notin\mathcal{S}$, $v$ can only be activated directly by its inverse neighbors $\mathcal{I}_v$ who are seed nodes in $\mathcal{S}$. Since each of such $v$'s inverse neighbors activates $v$ independently, the probability for all of them to fail to activate $v$ is $\prod_{w\in\mathcal{I}_v\cap\mathcal{S}}(1-p_{w,v})$. Consequently, the probability for $v$ to be activated is $1-\prod_{w\in\mathcal{I}_v\cap\mathcal{S}}(1-p_{w,v})$. Thus, for any node $v\in\mathcal{V}$, its one-hop activation probability is given by
\begin{equation}
\pi_1^{\mathcal{S}}(v)=
\begin{cases}
1, &\text{if}\ v\in\mathcal{S},\vspace{0.05in}\\
1-\prod_{w\in\mathcal{I}_v\cap\mathcal{S}}(1-p_{w,v}), &\text{otherwise}.
\end{cases}
\end{equation}

\begin{figure}[!t]
	\capstart
	\centering
	\includegraphics[height=1.3in]{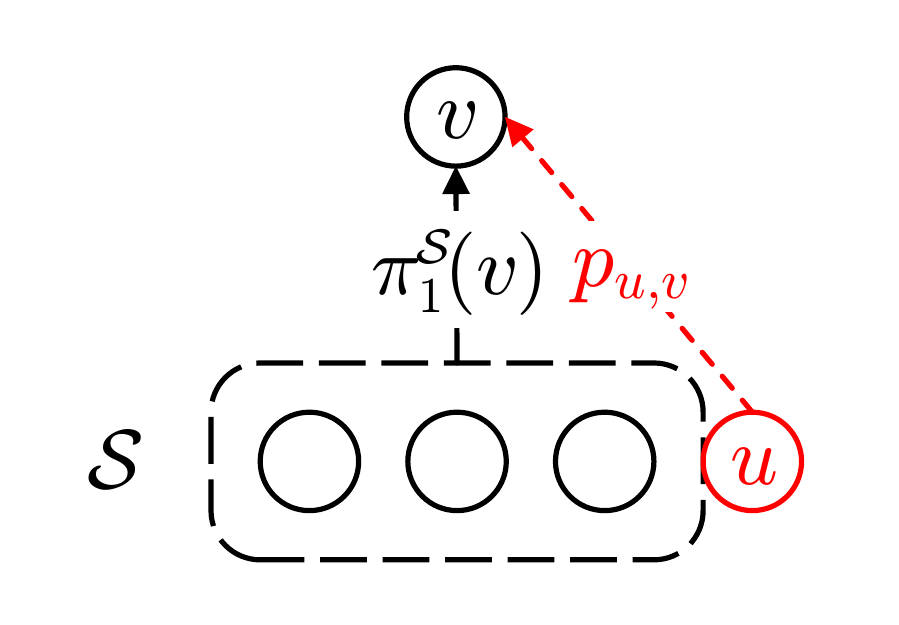}
	\vspace{-0.1in}
	\caption{The effect on $v\in\mathcal{N}_u$ by adding $u$ to the seed set $\mathcal{S}$.}\label{fig:H1}
	\vspace{-0.1in}
\end{figure}

Now, we show how to maintain $\pi_1^{\mathcal{S}}(v)$ when the seed set changes. Suppose that $\pi_1^{\mathcal{S}}(v)$ is known for every node $v\in\mathcal{V}$. If a new seed node $u$ is added to a seed set $\mathcal{S}$, it is clear that the activation probability of the new seed becomes $1$, i.e., $\pi_1^{\mathcal{S}\cup\{u\}}(u)=1$. In addition to $u$, only the one-hop activation probabilities of its neighbors are affected. So, we can incrementally update $\pi_1^{\mathcal{S}\cup\{u\}}(v)$ based on $\pi_1^{\mathcal{S}}(v)$ for each node $v\in\mathcal{N}_u$ (see \figurename~\ref{fig:H1}). The new activation probability $\pi_1^{\mathcal{S}\cup\{u\}}(v)$ is given by
\begin{align}
\pi_1^{\mathcal{S}\cup\{u\}}(v)
&=1-\!\!\!\!\!\!\prod_{w\in(\mathcal{I}_v\cap\mathcal{S})\cup\{u\}}\!\!\!\!\!\!(1-p_{w,v})\notag\\
&=1-\big(1-\pi_1^{\mathcal{S}}(v)\big)\cdot(1-p_{u,v}).
\label{eq:oneInc}
\end{align}
Algorithm~\ref{alg:oneHopInc} calculates the increment of one-hop influence spread $\sigma_1(\mathcal{S}\cup\{u\})-\sigma_1(\mathcal{S})$ efficiently by maintaining $\pi_1^{\mathcal{S}}(v)$ for every node $v$ based on the above equation.
\begin{algorithm}[!h]
	\capstart
	\label{alg:oneHopInc}
	\caption{$\bm{OneHopIncrement}(\mathcal{G},\mathcal{S}, u)$}
	$\pi_1^{\mathcal{S}\cup\{u\}}(u)\leftarrow 1$\;
	\For{each node $v\in \mathcal{N}_u\setminus \mathcal{S}$}{
		$\pi_1^{\mathcal{S}\cup\{u\}}(v)\leftarrow 1-\big(1-\pi_1^{\mathcal{S}}(v)\big)\cdot(1-p_{u,v})$\;
	}
	\Return $\sum_{v\in\{u\}\cup(\mathcal{N}_u\setminus\mathcal{S})}\big(\pi_1^{\mathcal{S}\cup\{u\}}(v)-\pi_1^{\mathcal{S}}(v)\big)$\;
\end{algorithm}

\begin{figure}[!t]
	\capstart
	\centering
	\includegraphics[height=1.3in]{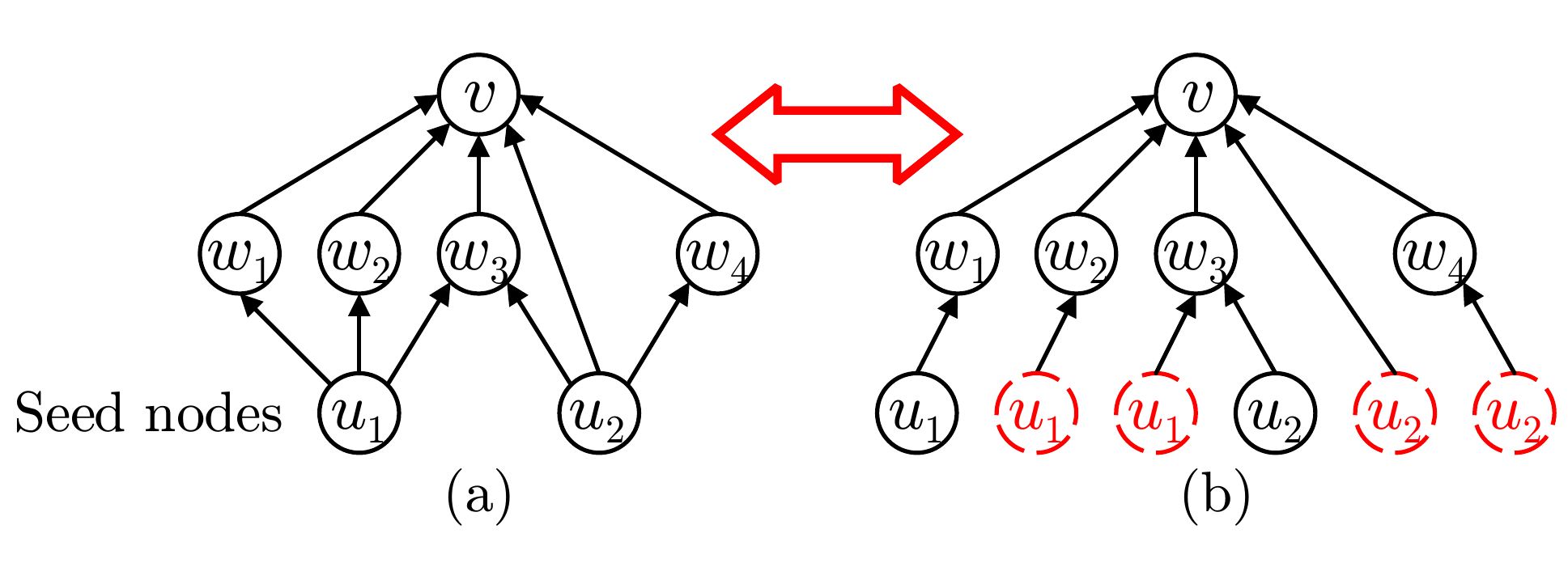}
	\vspace{-0.3in}
	\caption{An example of how a non-seed node $v$ is activated.}\label{fig:H2_example}
	\vspace{-0.1in}
\end{figure}

\textbf{Two Hops of Propagation:} To better approximate $\sigma(\mathcal{S})$, we next model two hops of propagation. As illustrated in \figurename~\ref{fig:H2_example}(a), with two hops of propagation, a non-seed node $v$ may be activated directly by a seed node $u_i$ or indirectly via a neighbor $w_j$ of a seed node $u_i$. In the former case, the probability for $v$ to be activated by $u_i$ is $p_{u_i,v}$, which can be rewritten as $p_{u_i,v}\cdot \pi_1^{\mathcal{S}}(u_i)$ since $\pi_1^{\mathcal{S}}(u_i)=1$. In the latter case, the probability for $v$ to be activated by $w_j$ is $p_{w_j,v}\cdot \pi_1^{\mathcal{S}}(w_j)$. Since the activation probability of each seed node $u_i\in\mathcal{S}$ is $1$, \figurename~\ref{fig:H2_example}(a) is equivalent to \figurename~\ref{fig:H2_example}(b) in which $v$ is activated independently by all of its inverse neighbors. As a result, the probability for $v$ to be activated is given by $1-\prod_{w\in\mathcal{I}_v}(1-p_{w,v}\cdot \pi_1^{\mathcal{S}}(w))$. Thus, for any node $v\in\mathcal{V}$, its two-hop activation probability is given by
\begin{equation}
\label{eq:twoHop}
\pi_2^{\mathcal{S}}(v)=
\begin{cases}
1, &\text{if}\ v\in\mathcal{S},\vspace{0.05in}\\
1-\prod_{w\in\mathcal{I}_v}\big(1-p_{w,v}\cdot \pi_1^{\mathcal{S}}(w)\big), &\text{otherwise}.
\end{cases}
\end{equation}

\begin{figure}[!t]
	\capstart
	\centering
	\includegraphics[height=1.3in]{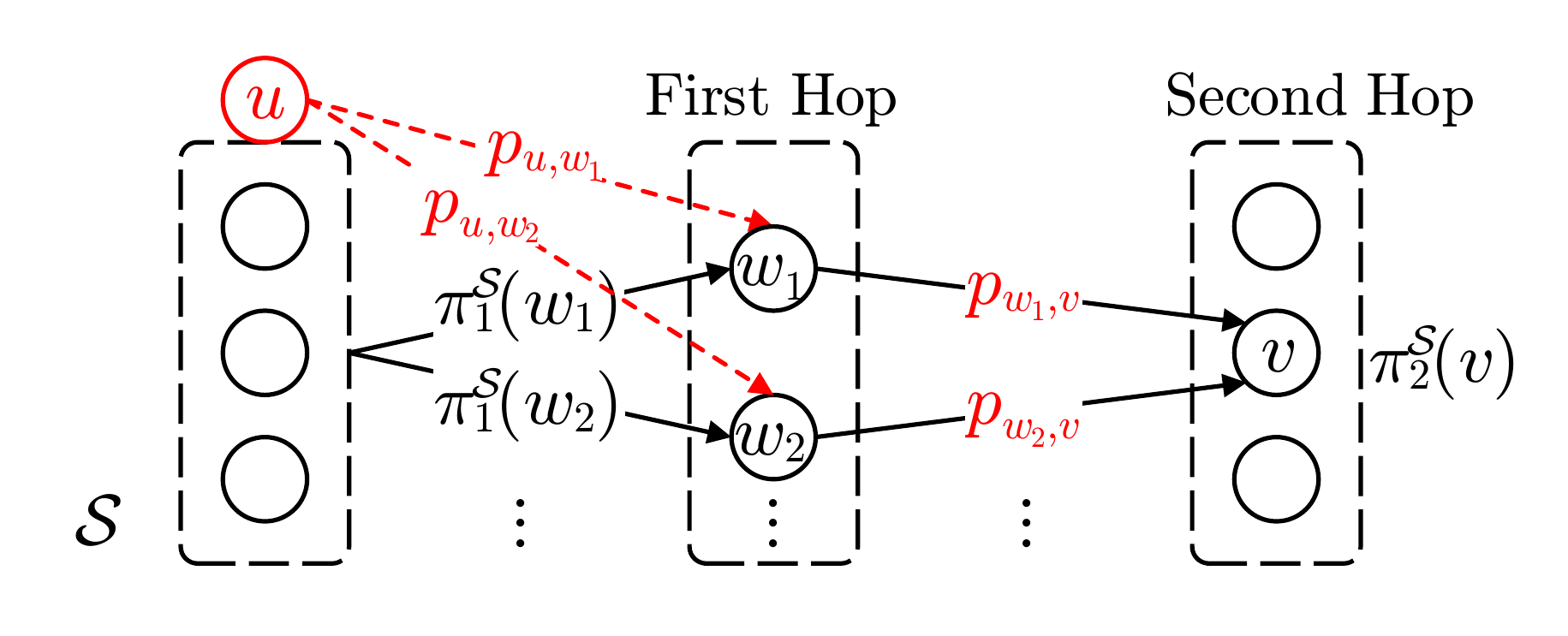}
	\vspace{-0.05in}
	\caption{The effect on $v\in\!\mathcal{N}_u^2$ by adding $u$ to the seed set $\mathcal{S}$.}\label{fig:H2}
	\vspace{-0.1in}
\end{figure}

According to the above equation, we can obtain $\pi_2^{\mathcal{S}}(v)$ based on the $\pi_1^{\mathcal{S}}(w)$'s of its inverse neighbors. When a new seed node $u$ is added, only the nodes within two hops of $u$ are affected. We denote these nodes by $\mathcal{N}_u^2=\mathcal{N}_u\cup\big(\bigcup_{w\in\mathcal{N}_u}\mathcal{N}_w\big)\setminus\{u\}$. \figurename~\ref{fig:H2} illustrates the effect of adding $u$ to $\mathcal{S}$ on the activation of a node $v$ in $\mathcal{N}_u^2$. Due to the independence among all two-hop activation paths as discussed above, we can incrementally update the two-hop activation probability by considering the outgoing edges from $u$ one at a time.

\begin{theorem}\label{theorem:twohop_incremental}
	The new two-hop activation probability $\pi_2^{\mathcal{S}\cup\{u\}}(v)$ after adding a seed $u$ to $\mathcal{S}$ can be computed by
	\begin{equation}
	\pi_2^{\mathcal{S}\cup\{u\}}(v) = 1-\big(1-\pi_2^{\mathcal{S}}(v)\big)\cdot\!\!\!\!\!\!\!\!\prod_{w\in(\mathcal{M}_{u,v}\cup\{u\})}\!\!\!\!\!\!\!\!\frac{1-p_{w,v}\cdot \pi_1^{\mathcal{S}\cup\{u\}}(w)}{1-p_{w,v}\cdot \pi_1^{\mathcal{S}}(w)},
	\end{equation}
	where $\mathcal{M}_{u,v}$ denotes the set of intermediate nodes connecting $u$ and $v$, i.e., $\mathcal{M}_{u,v}=\{w:(u,w)\in\mathcal{E}\text{ and } (w,v)\in\mathcal{E}\}$.
\end{theorem}

We leave the formal proofs of all theoretical results to the appendix.

Algorithm~\ref{alg:twoHopsInc} performs the updates on the activation probabilities of the nodes within two hops of $u$ when $u$ is added as a new seed to $\mathcal{S}$. Lines~\ref{line:twoHopsInc_setup1}--\ref{line:twoHopsInc_setup2} set the one-hop and two-hop activation probabilities of the new seed node $u$ to $1$. Lines~\ref{line:twoHopsInc_initialize1}--\ref{line:twoHopsInc_initialize2} initialize $\pi_2^{\mathcal{S}\cup\{u\}}(v)$ for all the nodes within two hops of $u$. Line~\ref{line:twoHopsInc_oneHop} computes the new one-hop activation probabilities for all of $u$'s neighbors as explained earlier. For each node $v\in\mathcal{N}_u^2\setminus\mathcal{S}$, lines~\ref{line:twoHopsInc_twoHop1}--\ref{line:twoHopsInc_twoHop2} calculate the new two-hop activation probability $\pi_2^{\mathcal{S}\cup\{u\}}(v)$ in an iterative manner according to Theorem~\ref{theorem:twohop_incremental}. In this way, we can save a huge amount of space for storing the intermediate nodes $\mathcal{M}_{u,v}$ for every pair of nodes $u$ and $v$. Finally, the algorithm returns the total increment of two-hop influence spread $\sigma_2(\mathcal{S}\cup\{u\})-\sigma_2(\mathcal{S})$.

\begin{algorithm}[!t]
	\capstart
	\label{alg:twoHopsInc}
	\caption{$\bm{TwoHopsIncrement}(\mathcal{G},\mathcal{S}, u)$}
	$\pi_1^{\mathcal{S}\cup\{u\}}(u)\leftarrow 1$\;\label{line:twoHopsInc_setup1}
	$\pi_2^{\mathcal{S}\cup\{u\}}(u)\leftarrow 1$\;\label{line:twoHopsInc_setup2}
	\For {each node $v\in\mathcal{N}_u^2\setminus\mathcal{S}$}{\label{line:twoHopsInc_initialize1}
		$\pi_2^{\mathcal{S}\cup\{u\}}(v)\leftarrow \pi_2^{\mathcal{S}}(v)$\;\label{line:twoHopsInc_initialize2}
	}
	\For{each node $w\in \mathcal{N}_u\setminus \mathcal{S}$}{
		$\pi_1^{\mathcal{S}\cup\{u\}}(w)\leftarrow 1-\big(1-\pi_1^{\mathcal{S}}(w)\big)\cdot(1-p_{u,w})$\;\label{line:twoHopsInc_oneHop}
		$\pi_2^{\mathcal{S}\cup\{u\}}(w)\leftarrow 1-\big(1-\pi_2^{\mathcal{S}\cup\{u\}}(w)\big)\cdot\frac{1-p_{u,w}\cdot \pi_1^{\mathcal{S}\cup\{u\}}(u)}{1-p_{u,w}\cdot \pi_1^{\mathcal{S}}(u)}$\;\label{line:twoHopsInc_twoHop1}
		\For{each node $v\in\mathcal{N}_w\setminus \mathcal{S}$}{
			$\!\!\!\pi_2^{\mathcal{S}\cup\{u\}}(v)\leftarrow 1-\big(1-\pi_2^{\mathcal{S}\cup\{u\}}\!(v)\big)\cdot\frac{1-p_{w,v}\cdot \pi_1^{\mathcal{S}\cup\{u\}}(w)}{1-p_{w,v}\cdot \pi_1^{\mathcal{S}}(w)}$\;\label{line:twoHopsInc_twoHop2}
		}
	}
	\Return $\sum_{v\in \{u\}\cup(\mathcal{N}_u^2\setminus\mathcal{S})}\big(\pi_2^{\mathcal{S}\cup\{u\}}(v)-\pi_2^{\mathcal{S}}(v)\big)$\;
\end{algorithm}

\subsection{Complexity}\label{subsec:complexity}
We shall refer to the greedy heuristic (Algorithm~\ref{alg:greedy}) as the OneHop and TwoHop algorithms respectively when the influence spread is approximated by one hop and two hops of propagation. 

\textbf{Time Complexity:} The time complexity of Algorithm~\ref{alg:oneHopInc} is $O(1+|\mathcal{N}_u|)$. Thus, the time complexity of selecting one seed in the OneHop algorithm is $O\big(\sum_{u\in\mathcal{V}}(1+|\mathcal{N}_u|)\big)=O\big(|\mathcal{V}|+|\mathcal{E}|\big)$. Therefore, the total time complexity of OneHop is $O\big(k(|\mathcal{V}|+|\mathcal{E}|)\big)$.
The time complexity of Algorithm~\ref{alg:twoHopsInc} is $O\big(1+|\mathcal{N}_u|+\sum_{w\in\mathcal{N}_u}|\mathcal{N}_w|\big)$. Therefore, the time complexity of selecting one seed in the TwoHop algorithm is $O\big(\sum_{u\in\mathcal{V}}(1+|\mathcal{N}_u|+\sum_{w\in\mathcal{N}_u}|\mathcal{N}_w|)\big)=O\big(|\mathcal{V}|+|\mathcal{E}|+\sum_{u\in\mathcal{V}}\sum_{w\in\mathcal{N}_u}|\mathcal{N}_w|\big)=O\big(|\mathcal{V}|+|\mathcal{E}|+\sum_{w\in\mathcal{V}}(|\mathcal{I}_w|\cdot|\mathcal{N}_w|)\big)$. Thus, the total time complexity of TwoHop is $O\Big(k\big(|\mathcal{V}|+|\mathcal{E}|+\sum_{w\in\mathcal{V}}(|\mathcal{I}_w|\cdot|\mathcal{N}_w|)\big)\Big)$.

\textbf{Space Complexity:} Besides the space used to store the graph, the OneHop algorithm only requires $O(|\mathcal{V}|)$ space to store the one-hop activation probability $\pi_1^{\mathcal{S}}(v)$ for every $v\in\mathcal{V}$ before and after a new seed is added. Similarly, the TwoHop algorithm requires $O(|\mathcal{V}|)$ space to store the one-hop and two-hop activation probabilities $\pi_1^{\mathcal{S}}(v)$ and $\pi_2^{\mathcal{S}}(v)$ for computing the influence increment. Thus, the space complexities of the OneHop and TwoHop algorithms are both $O(|\mathcal{V}|)$.

\subsection{Further Improvement on Efficiency}\label{subsec:improvement}

Note that selecting the first seed in Algorithm~\ref{alg:greedy} requires calculating the influence spread $\sigma(\{v\})$ for every node $v$ even when the CELF technique \cite{Leskovec_CELF_2007} is adopted. To avoid such computation, we develop an upper bound on $\sigma(\{v\})$ when hop-based influence estimation is applied.

\begin{theorem}\label{theorem:ubound}
	For each node $v\in\mathcal{V}$, the $h$-hop influence spread $\sigma_h(\{v\})$ satisfies
	\begin{equation}\label{eq:ubound1}
	\sigma_h(\{v\}) \leq 1+\sum_{w\in\mathcal{N}_v}\big(p_{v,w}\cdot \sigma_{h-1}(\{w\})\big).
	\end{equation}
	Furthermore, let $\hat{\sigma}_{0}(\{v\})=\sigma_{0}(\{v\})=1$ and $\hat{\sigma}_h(\{v\}) = 1+\sum_{w\in\mathcal{N}_v}\big(p_{v,w}\cdot \hat{\sigma}_{h-1}(\{w\})\big)$, then
	\begin{equation}\label{eq:ubound2}
	\sigma_h(\{v\}) \leq \hat{\sigma}_h(\{v\}).
	\end{equation}
\end{theorem}

Note that when $h=1$, the upper bound $\hat{\sigma}_{1}(\{v\})=1+\sum_{w\in\mathcal{N}_v}p_{v,w}$ is the exact $1$-hop influence spread of a single seed $\{v\}$, i.e., $\hat{\sigma}_{1}(\{v\})={\sigma}_{1}(\{v\})$. Computing $\hat{\sigma}_1(\{v\})$ for a node $v$ has a time complexity of $O(1+|\mathcal{N}_v|)$. Thus, it takes a time complexity of $O(\sum_{v\in\mathcal{V}}(1+|\mathcal{N}_v|))=O(|\mathcal{V}|+|\mathcal{E}|)$ to calculate $\hat{\sigma}_1(\{v\})$ for all nodes $v\in\mathcal{V}$. The time complexity for computing the upper bound $\hat{\sigma}_2(\{v\})$ given in Theorem~\ref{theorem:ubound} is $O(1+|\mathcal{N}_v|)$ after obtaining $\hat{\sigma}_1(\{w\})$ for all nodes $w\in\mathcal{V}$. Thus, the total time complexity for calculating $\hat{\sigma}_2(\{v\})$ for all nodes $v\in\mathcal{V}$ is $O(|\mathcal{V}|+|\mathcal{E}|)+O(\sum_{v\in\mathcal{V}}(1+|\mathcal{N}_v|))=O(|\mathcal{V}|+|\mathcal{E}|)+O(|\mathcal{V}|+|\mathcal{E}|)=O(|\mathcal{V}|+|\mathcal{E}|)$, which is much lower than the time complexity for computing the exact two-hop influence spread ${\sigma}_{2}(\{v\})$ for all nodes $v \in \mathcal{V}$ using Algorithm~\ref{alg:twoHopsInc} which is $O\big(|\mathcal{V}|+|\mathcal{E}|+\sum_{w\in\mathcal{V}}(|\mathcal{I}_w|\cdot|\mathcal{N}_w|)\big)$. On obtaining the upper bounds $\hat{\sigma}_h(\{v\})$, the CELF technique described in Section~\ref{subsec:greedy} can then be applied to the first iteration of Algorithm~\ref{alg:greedy} so that only the influence spreads of a subset of nodes in $\mathcal{V}$ need to be calculated for selecting the first seed. We shall show in Section~\ref{subsec:results} that the upper bounding approach can dramatically reduce the running time of the two-hop method.

\subsection{Theoretical Analysis}\label{subsec:analysis}

In this section, we carry out theoretical analysis for our hop-based algorithms. We first show that the influence spread within $h$ hops of propagation is submodular and monotone.

\begin{theorem}\label{theorem:hop_influence}
	For any $h\geq 1$, the influence spread produced within $h$ hops of propagation is submodular and monotone under the IC model.
\end{theorem}

Let $\mathcal{S}^*$ denote the optimal seed set for maximizing the actual influence without any hop limit, i.e., $\sigma(\mathcal{S}^*)=\max_{|\mathcal{S}|=k}\sigma(\mathcal{S})$. Then, we can derive the following guarantees for our hop-based methods.  

\begin{theorem}\label{theorem:ratio_guarantee}
	Under the IC model, if $\sigma_h(\mathcal{S})/\sigma(\mathcal{S})\geq\alpha$ for any seed set $|\mathcal{S}|=k$, the solution $\mathcal{S}_h$ returned by the greedy heuristic (Algorithm~\ref{alg:greedy}) with hop-based influence estimation satisfies
	\begin{equation}
	\sigma(\mathcal{S}_h)\geq \big((1-1/e)\alpha\big)\cdot\sigma(\mathcal{S}^*).
	\end{equation}
\end{theorem}

Theorem~\ref{theorem:ratio_guarantee} indicates that if the ratio $\sigma_h(\mathcal{S})/\sigma(\mathcal{S})$ is lower bounded by $\alpha$, the hop-based methods can provide a multiplicative guarantee of $\big((1-1/e)\alpha\big)$. Next, we derive a lower bound on the ratio $\sigma_h(\mathcal{S})/\sigma(\mathcal{S})$ in the class of scale free random graphs which are commonly used to model OSNs \cite{Barabasi_1999,Li_advertising_2012}. The degree distribution of a scale-free (undirected) graph follows a power law. That is, the probability of a node having degree $d$ is $P_0(d)=\frac{d^{-\gamma}}{\sum_{d_i=1}^{\infty}d_i^{-\gamma}}$, where $\gamma$ is a given power scale parameter whose typical value is in the range of $2\leq \gamma \leq 3$. We first analyze the expected number of nodes activated within one hop of propagation, which gives a lower bound on $\sigma_h(\mathcal{S})$ for any $h \geq 1$ since $\sigma_h(\mathcal{S})$ increases with $h$. Next, we derive an upper bound on the expected number of nodes activated $\sigma(\mathcal{S})$. Using the lower bound on $\sigma_h(\mathcal{S})$ and upper bound on $\sigma(\mathcal{S})$, we can derive a lower bound $\alpha$ on the ratio $\sigma_h(\mathcal{S})/\sigma(\mathcal{S})$.
\begin{theorem}\label{theorem:ratio}
	For scale free random graphs with propagation probability $p_{u,v}=p$ for every edge $(u,v)\in\mathcal{E}$ and any seed set $\mathcal{S}$ and any hop of $h\geq1$, we have
	\begin{equation}\label{eq:expected_influence_lower_bound}
	\frac{\mathbb{E}[\sigma_h(\mathcal{S})]}{\mathbb{E}[\sigma(\mathcal{S})]}\geq\frac{1-(1-k/|\mathcal{V}|)(1-pk/|\mathcal{V}|)}{1-(1-k/|\mathcal{V}|)P_0(1)(1-pA)},
	\end{equation}
	where $A=1-\big(1-\frac{k}{|\mathcal{V}|}\big)P_1(1)$ and $P_1(d)=\frac{d^{1-\gamma}}{\sum_{d_i=1}^{\infty}d_i^{1-\gamma}}$.
\end{theorem}

\begin{figure}[t]
	\capstart
	\centering
	\includegraphics[width=0.75\linewidth]{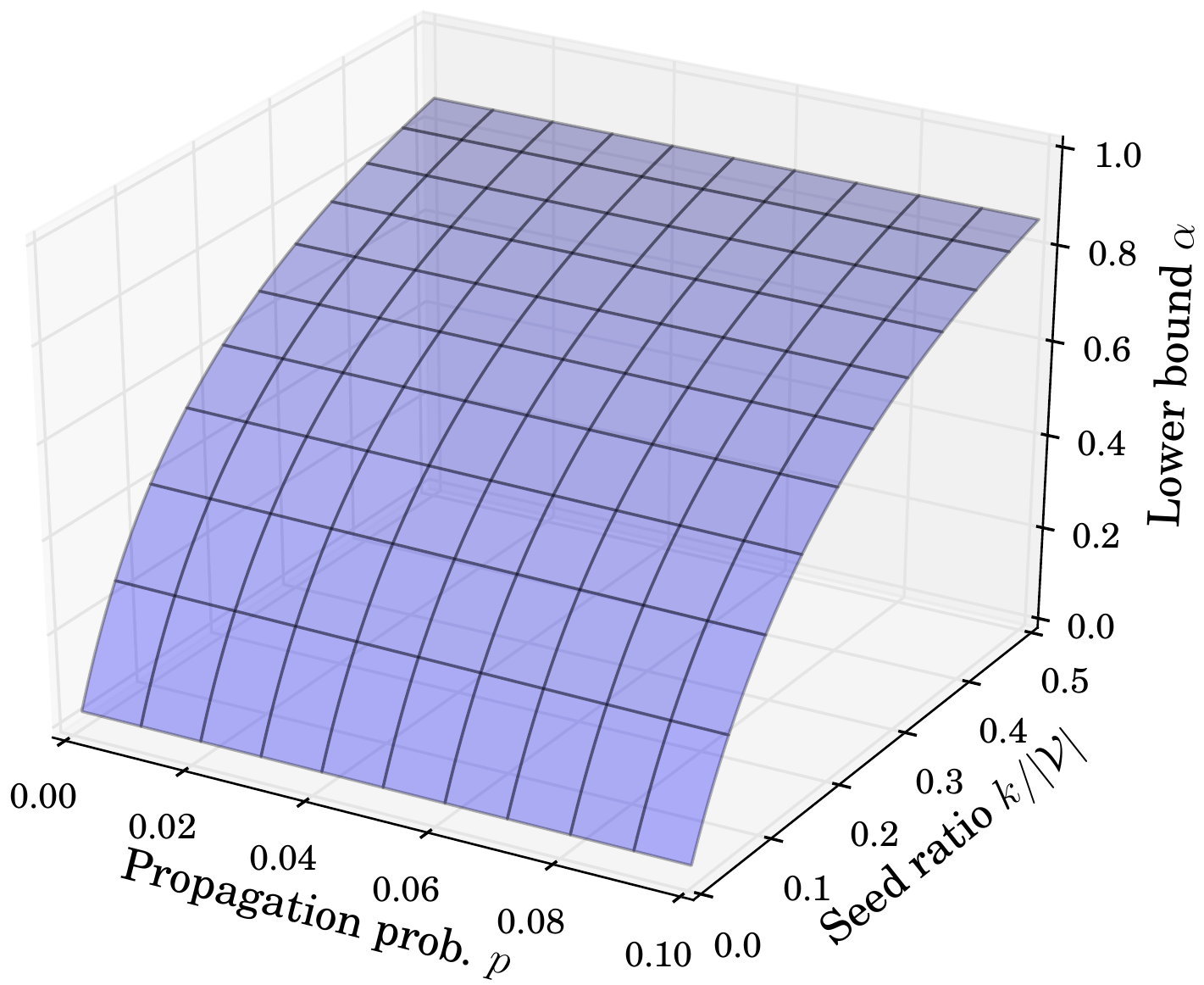}\label{lower_bound}
	\caption{Lower bound $\alpha$ for different propagation probabilities and seed ratios ($\gamma=3$).}\label{fig:lower_bound}
	\vspace{-0.1in}
\end{figure}

\figurename~\ref{fig:lower_bound} shows the lower bound derived in (\ref{eq:expected_influence_lower_bound}) when varying the propagation probability $p$ from $0$ to $0.1$ and the seed ratio $k/|\mathcal{V}|$ from $0$ to $0.5$. We can see that the lower bound generally increases with both the seed ratio and propagation probability.

\subsection{Extension to Linear Threshold Model}\label{subsec:LT}
The Linear Threshold (LT) model is another basic diffusion model described in \cite{Kempe_maxInfluence_2003}, which is also widely used \cite{Chen_LDAG_2010,Tang_IMM_2015,Tang_reverse_2014}. In the LT model, each directed edge $(u,v)$ is associated with a weight $b_{u,v}$, and the total weight of all the incoming edges to each node does not exceed $1$. In the diffusion process, each node randomly chooses a threshold between $0$ and $1$. The diffusion starts from an initial seed set $\mathcal{S}$. A non-seed node becomes activated only when the aggregate weight of the incoming edges from its inverse neighbors that have been activated reaches its threshold.

Our proposed hop-based algorithms can be easily extended to work with the LT model. For one hop of propagation, if $v\in\mathcal{S}$, $\pi_1^{\mathcal{S}}(v)=1$; otherwise, $\pi_1^{\mathcal{S}}(v)=\sum_{u\in\mathcal{I}_v\cap\mathcal{S}}b_{u,v}$. When adding a new seed $u$ to a seed set $\mathcal{S}$, for each neighbor $v$ of $u$ where $v\notin\mathcal{S}$, the increment of $v$'s activation probability is $\pi_1^{\mathcal{S}\cup\{u\}}(v)-\pi_1^{\mathcal{S}}(v)=b_{u,v}$. Thus, the total increment of influence spread by adding $u$ to $\mathcal{S}$ is $\sigma_1(\mathcal{S}\cup\{u\})-\sigma_1(\mathcal{S})=\sum_{v\in\mathcal{N}_u\setminus\mathcal{S}}b_{u,v}$. For two hops of propagation, all the paths from seed nodes to a non-seed node within two hops are acyclic. The increment of influence spread by adding a new seed $u$ to a seed set $\mathcal{S}$ consists of three parts: 1) $1-\pi_2^{\mathcal{S}}(u)$ from node $u$ itself; 2) $\big(1-\pi_1^{\mathcal{S}}(u)\big)\cdot\sum_{v\in\mathcal{N}_u\setminus\mathcal{S}}b_{u,v}$ from $u$'s non-seed neighbors; 3) $\sum_{v\in\mathcal{N}_u\setminus\mathcal{S}}\big(b_{u,v}\cdot\sum_{w\in\mathcal{N}_v\setminus(\mathcal{S}\cup\{u\})}b_{v,w}\big)$ from $u$'s neighbors' neighbors which are not seeds. Thus, the total increment of influence spread by adding a new seed $u$ to a seed set $\mathcal{S}$ is $\sigma_2(\mathcal{S}\cup\{u\})-\sigma_2(\mathcal{S})=1-\pi_2^{\mathcal{S}}(u)+\big(1-\pi_1^{\mathcal{S}}(u)\big)\cdot\sum_{v\in\mathcal{N}_u\setminus\mathcal{S}}b_{u,v}+\sum_{v\in\mathcal{N}_u\setminus\mathcal{S}}\big(b_{u,v}\cdot\sum_{w\in\mathcal{N}_v\setminus(\mathcal{S}\cup\{u\})}b_{v,w}\big)$. It is easy to show that the influence spread within a fixed number of hops under the LT model is submodular and monotone as well. Thus, our analysis on the hop-based algorithms can also be applied to the LT model.

\section{Evaluation}\label{sec:evaluation}

\subsection{Experimental Setup}\label{subsec:setup}

\textbf{Datasets.} We use several real OSN datasets in our experiments \cite{NetHEPT,Kwak_twitter_2010,snapnets}. Due to space limitations, we just show the results for three representative datasets: NetHEPT (15K nodes and 32K edges), LiveJournal (5M nodes and 69M edges) and Twitter (42M nodes and 1.5B edges).  The results for other datasets are similar.

\begin{figure*}[!t]
	\capstart
	\centering
	\includegraphics[width=1.0\linewidth]{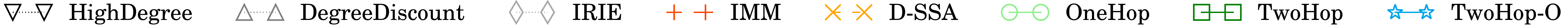}\vspace{-0.15in}\\
	\subfloat[NetHEPT]{\includegraphics[width=0.31\linewidth]{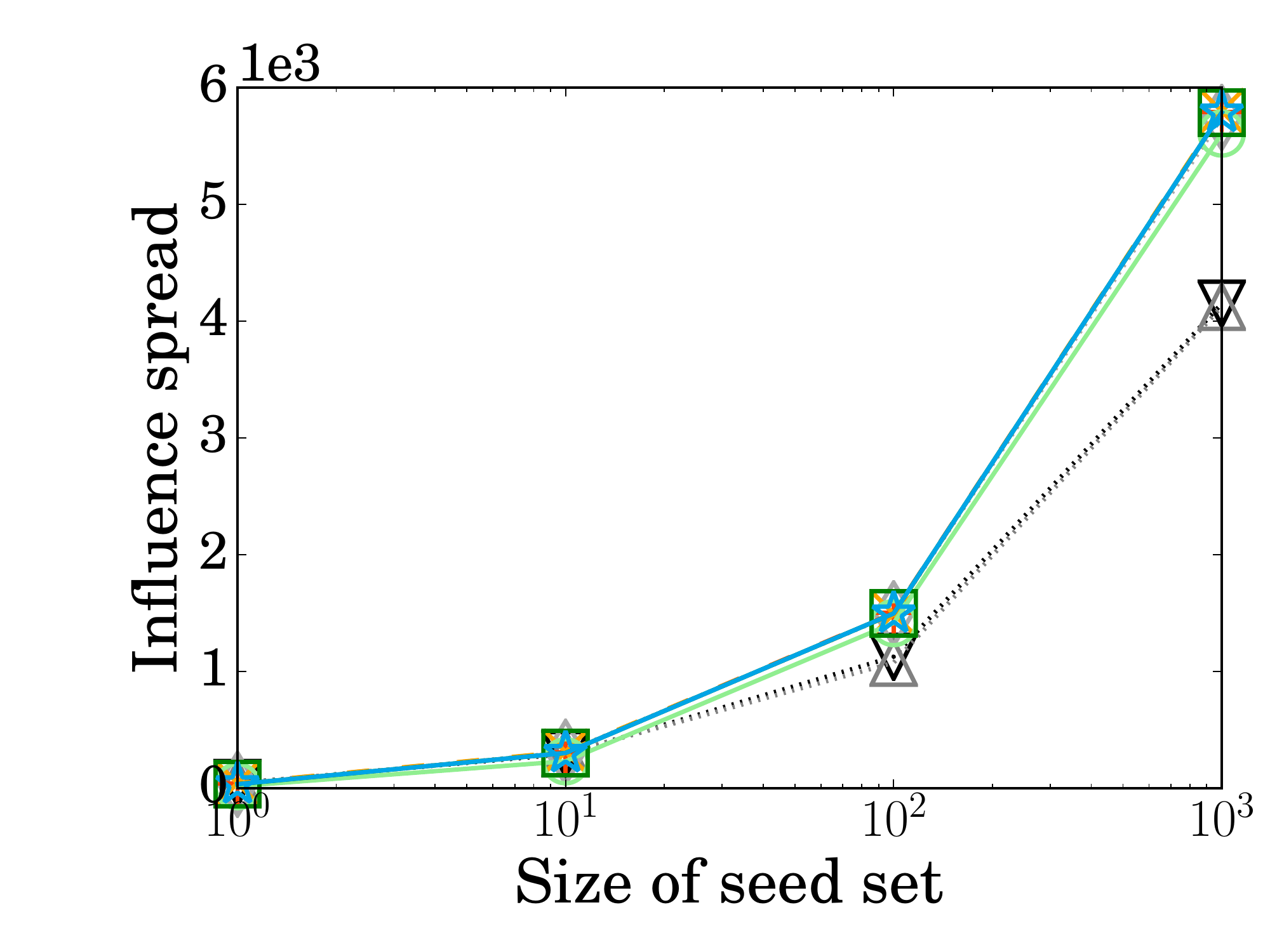}\label{subfig:NetHEPT_WC_inf}}\hfill
	\subfloat[LiveJournal]{\includegraphics[width=0.31\linewidth]{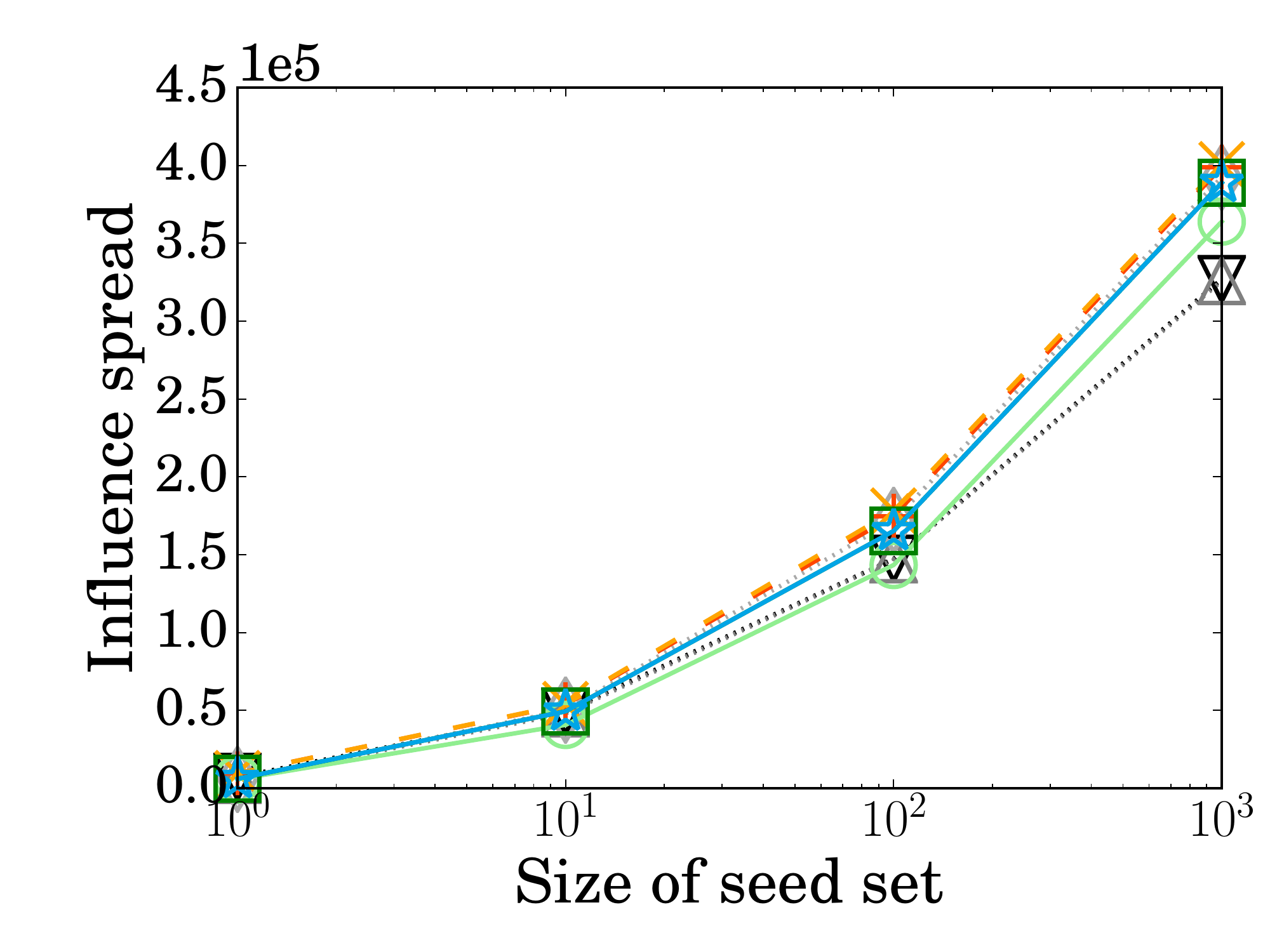}\label{subfig:livejournal_WC_inf}}\hfill
	\subfloat[Twitter (IRIE cannot run)]{\includegraphics[width=0.31\linewidth]{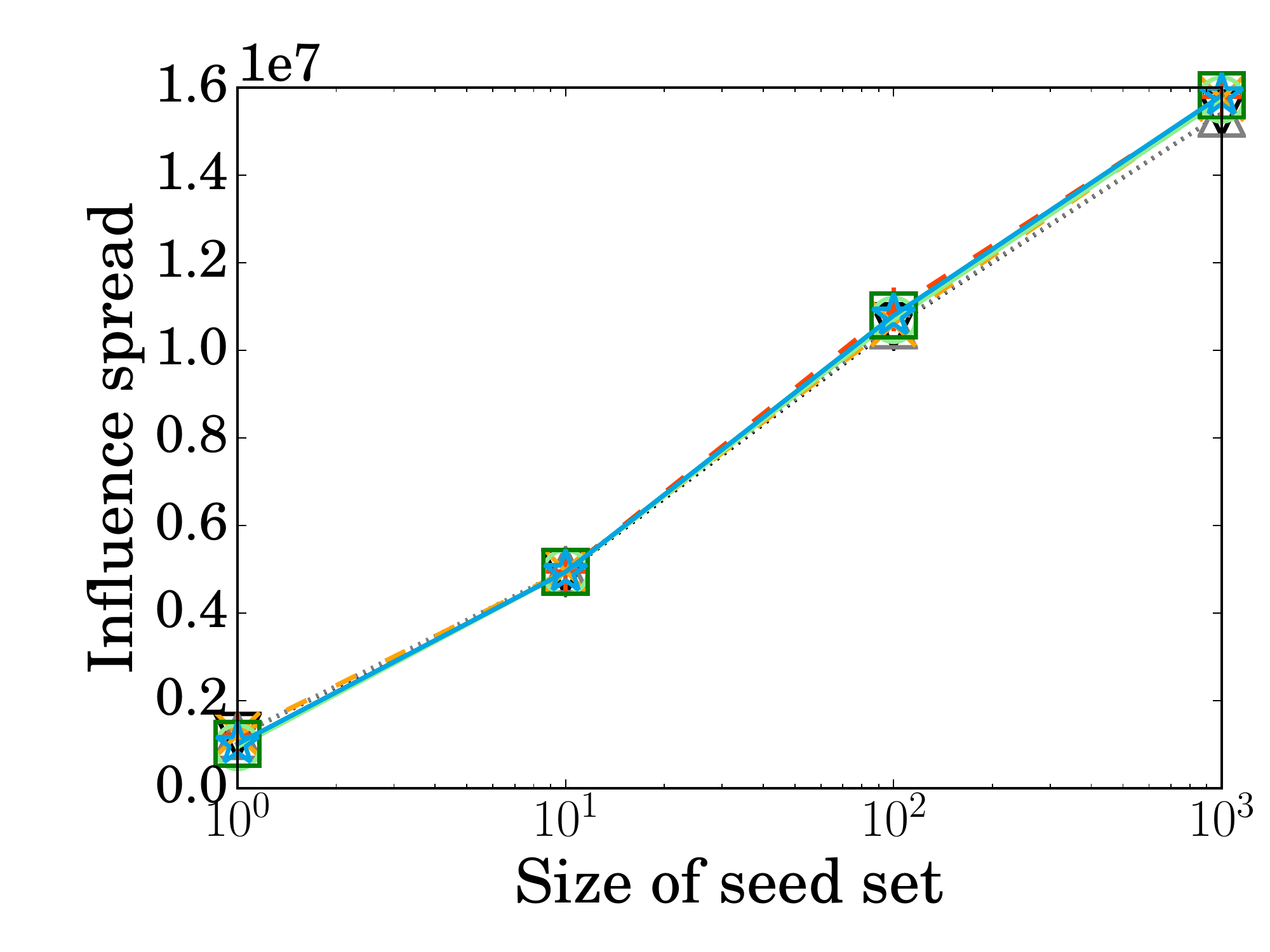}\label{subfig:big_twitter_WC_inf}}
	\caption{Influence spread on various graphs under the WC model.}\label{fig:influence_WC}
\end{figure*}

\begin{figure*}[!t]
	\capstart
	\centering
	\includegraphics[width=1.0\linewidth]{legend}\vspace{-0.15in}\\
	\subfloat[NetHEPT]{\includegraphics[width=0.31\linewidth]{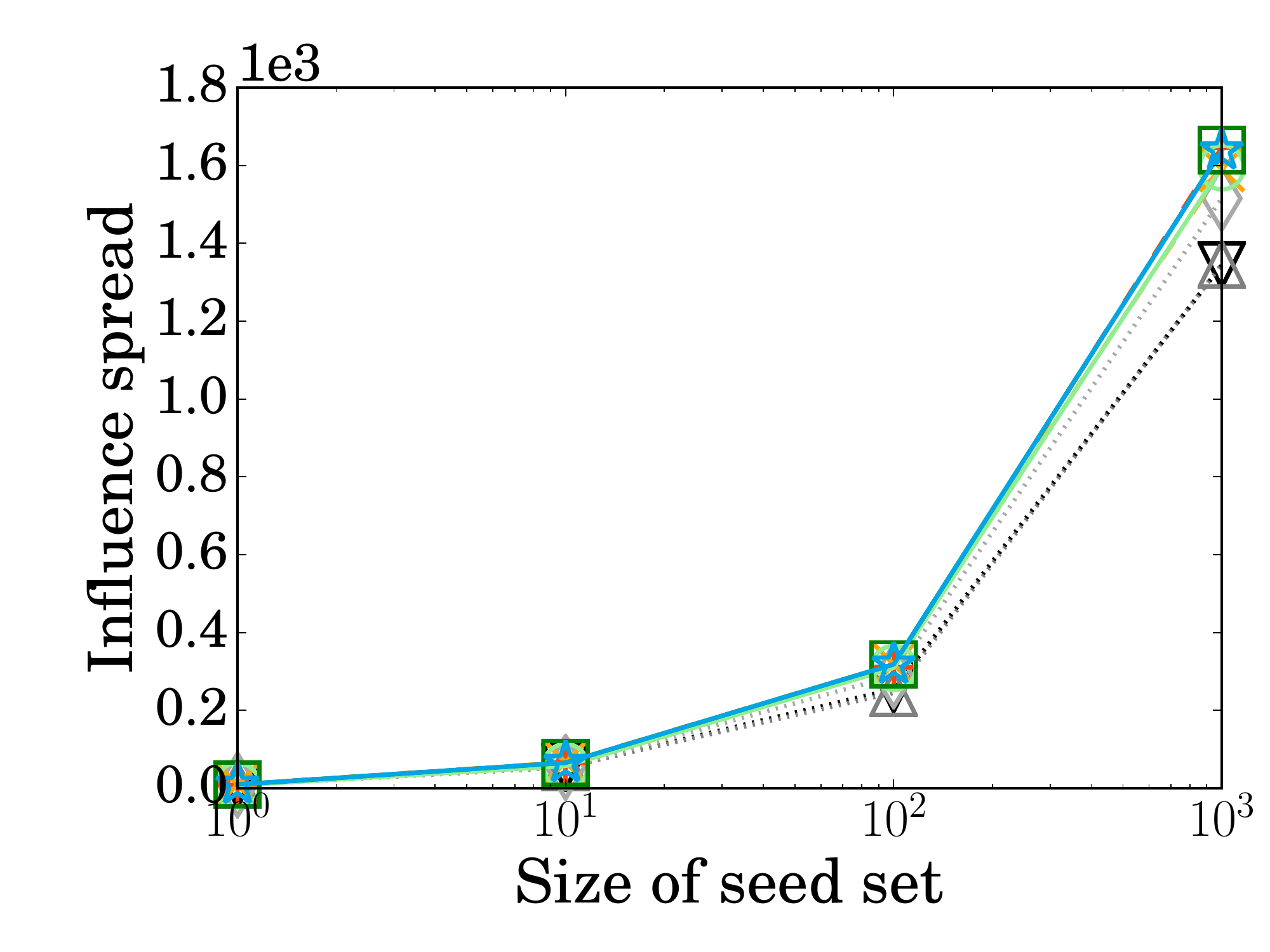}\label{subfig:NetHEPT_TR_inf}}\hfill
	\subfloat[LiveJournal (IMM and D-SSA cannot run)
	]{\includegraphics[width=0.31\linewidth]{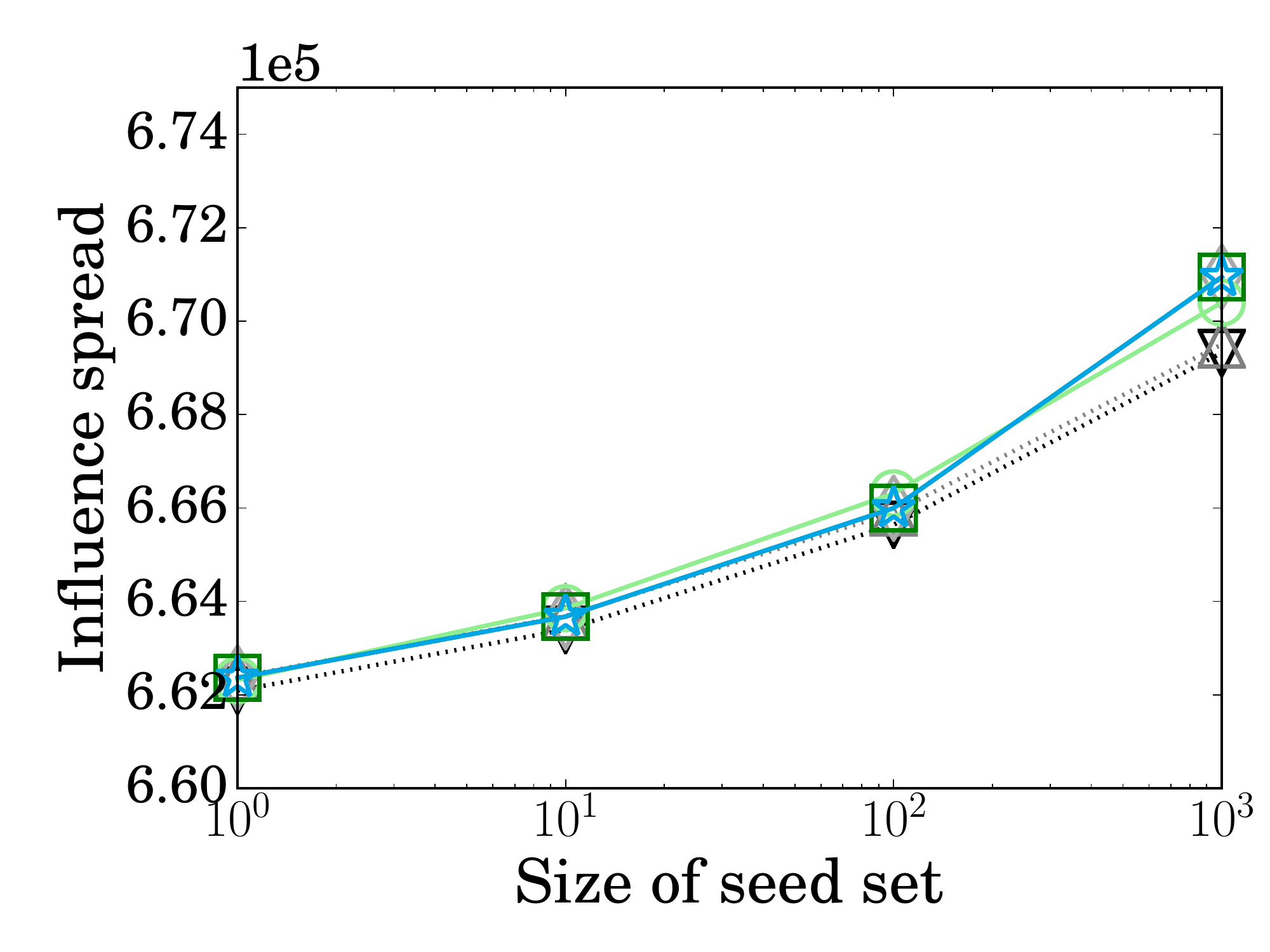}\label{subfig:livejournal_TR_inf}}\hfill
	\subfloat[Twitter (IRIE, IMM and D-SSA cannot run)]{\includegraphics[width=0.31\linewidth]{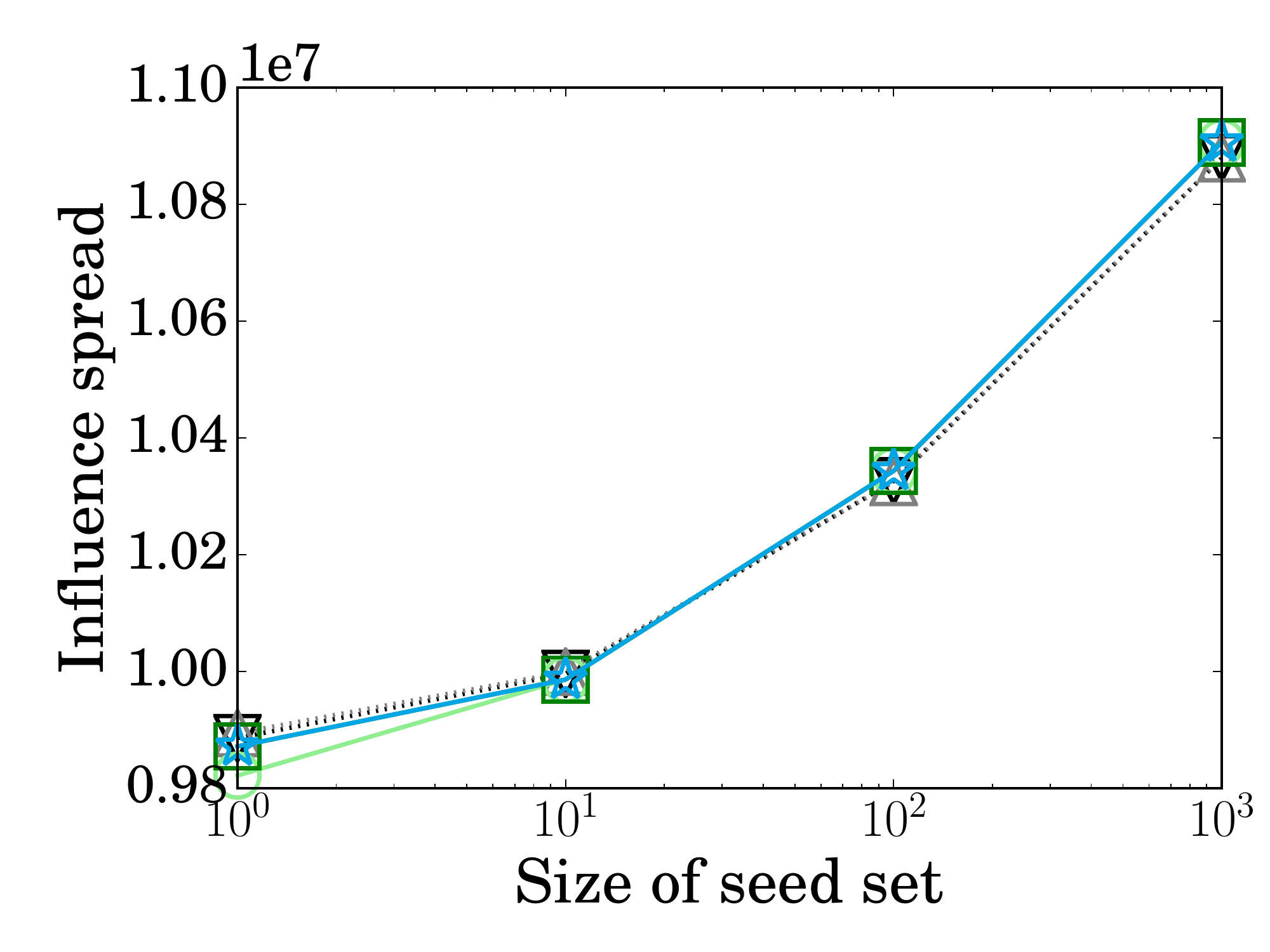}\label{subfig:big_twitter_TR_inf}}
	\caption{Influence spread on various graphs under the TRIVALENCY model.}\label{fig:influence_TR}
	\vspace{-0.1in}
\end{figure*}

\textbf{Algorithms.} We compare our OneHop, TwoHop and TwoHop-O (without the upper bounding technique described in Section~\ref{subsec:improvement}) algorithms with the following state-of-the-art algorithms.
\begin{itemize}
	\item HighDegree: Select the $k$ nodes with highest degrees \cite{Kempe_maxInfluence_2003}.
	\item DegreeDiscount: The degree discount heuristic was developed by \cite{Chen_degreeDiscount_2009}.
	\item IRIE: IRIE \cite{Jung_IRIE_2012} is a state-of-the-art heuristic. We set $\alpha=0.7$ and $\theta=1/320$ respectively as suggested in \cite{Jung_IRIE_2012}.
	\item IMM: IMM \cite{Tang_IMM_2015} is one of the most advanced sampling-based methods that can provide a $(1-1/e-\epsilon)$-approximation guarantee with probability at least $1-\delta$. 
	\item D-SSA: D-SSA \cite{Nguyen_DSSA_2016} aims to further reduce the number of samples generated compared to IMM while providing the same approximation guarantee. We set $\epsilon=0.1$ and $\delta=1/|\mathcal{V}|$ for both IMM and D-SSA according to the default setting in \cite{Nguyen_DSSA_2016}.
\end{itemize}

\textbf{Parameter Settings.} For the IC model, we set the propagation probability via the following two models.
\begin{itemize}
	\item WC model \cite{Chen_degreeDiscount_2009,Kempe_maxInfluence_2003,Nguyen_DSSA_2016,Tang_IMM_2015}: $p_{u,v}$ of each edge $(u,v)$ is set to the reciprocal of $v$'s in-degree, i.e., $p_{u,v}=1/|\mathcal{I}_v|$.
	\item TRIVALENCY model \cite{Chen_MIA_2010,Jung_IRIE_2012}: $p_{u,v}$ of each edge $(u,v)$ is set by choosing a probability from the set $\{0.1,0.01,0.001\}$ at random.
\end{itemize}

To evaluate the seed sets returned by different algorithms, we estimate the influence spread of each seed set by taking the average measurement of $10,000$ Monte-Carlo simulations. The algorithms are all implemented in C++ and the experiments are carried out on a machine with an Intel Xeon E5-2695 2.4GHz CPU and 64GB memory. We limit the running time of each algorithm up to $100$ hours ($3.6 \times 10^5$ seconds).

\subsection{Results}\label{subsec:results}

\textbf{Influence Spread:} \figuresname~\ref{fig:influence_WC} and \ref{fig:influence_TR} show the influence spread produced by different algorithms on various graphs when the size of seed set is set to $k=1,10,100,1000$ under the WC and TRIVALENCY models respectively. Due to out-of-memory reasons and prohibitively long computation times, IRIE \textit{failed} to produce results on the Twitter dataset under both WC and TRIVALENCY models, while both IMM and D-SSA \textit{failed} on the LiveJournal and Twitter datasets under the TRIVALENCY model. 
From the results obtained, we can make the following observations. Our OneHop, TwoHop and TwoHop-O methods usually generate influence spread as high as that by the IMM and D-SSA methods which can provide the state-of-the-art $(1-1/e-\epsilon)$-approximation guarantee. Our methods remarkably outperform both the HighDegree and DegreeDiscount heuristics (by up to $40\%$) on the NetHEPT dataset under both WC and TRIVALENCY models (\figuresname~\ref{subfig:NetHEPT_WC_inf}, \ref{subfig:NetHEPT_TR_inf}) and on the LiveJournal dataset under the WC model (\figurename~\ref{subfig:livejournal_WC_inf}). These observations demonstrate the effectiveness of our hop-based methods.

\begin{figure*}[!t]
	\capstart
	\centering
	\includegraphics[width=1.0\linewidth]{legend}\vspace{-0.15in}\\
	\subfloat[NetHEPT]{\includegraphics[width=0.31\linewidth]{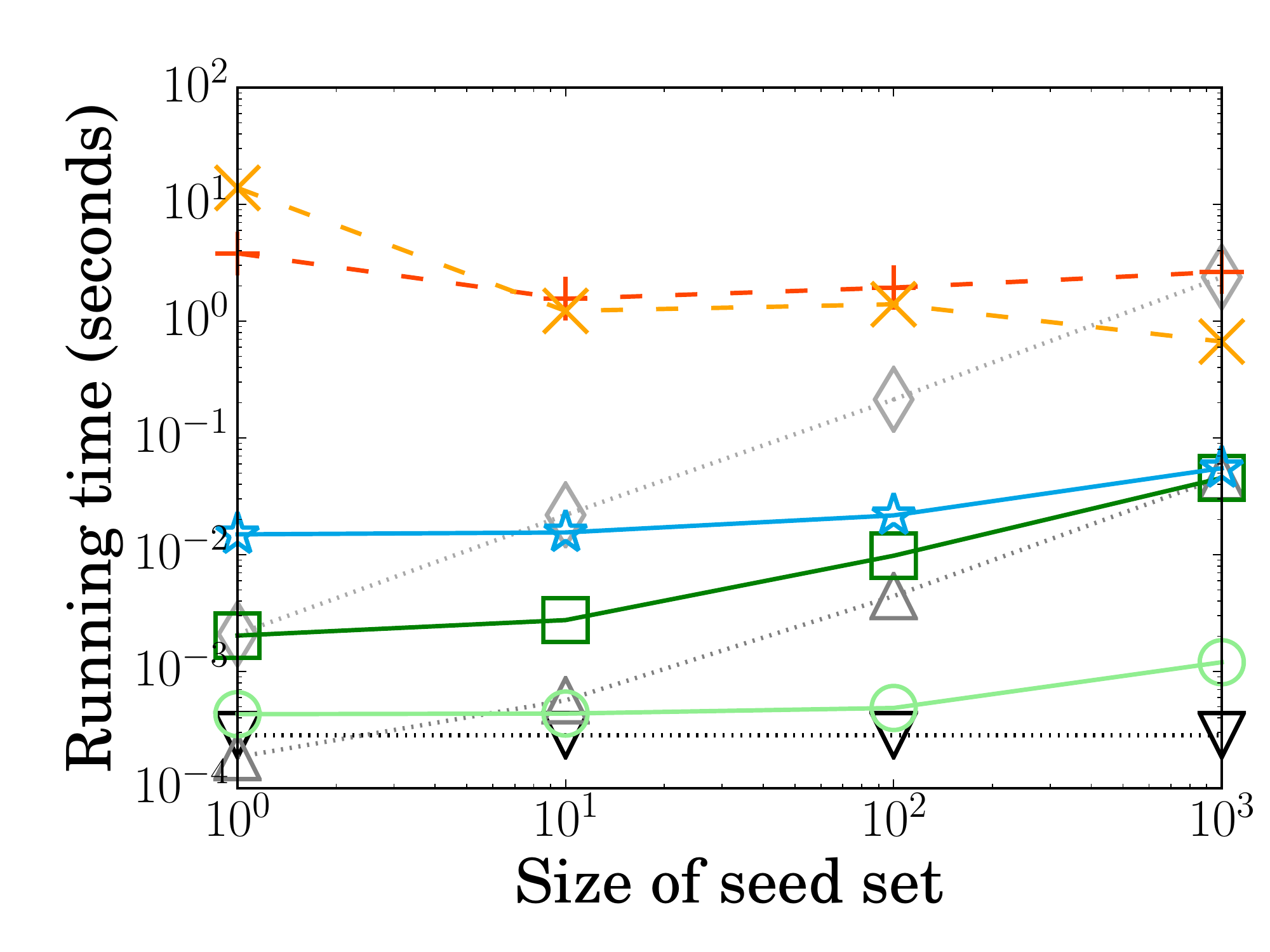}\label{subfig:NetHEPT_WC_time}}\hfill
	\subfloat[LiveJournal]{\includegraphics[width=0.31\linewidth]{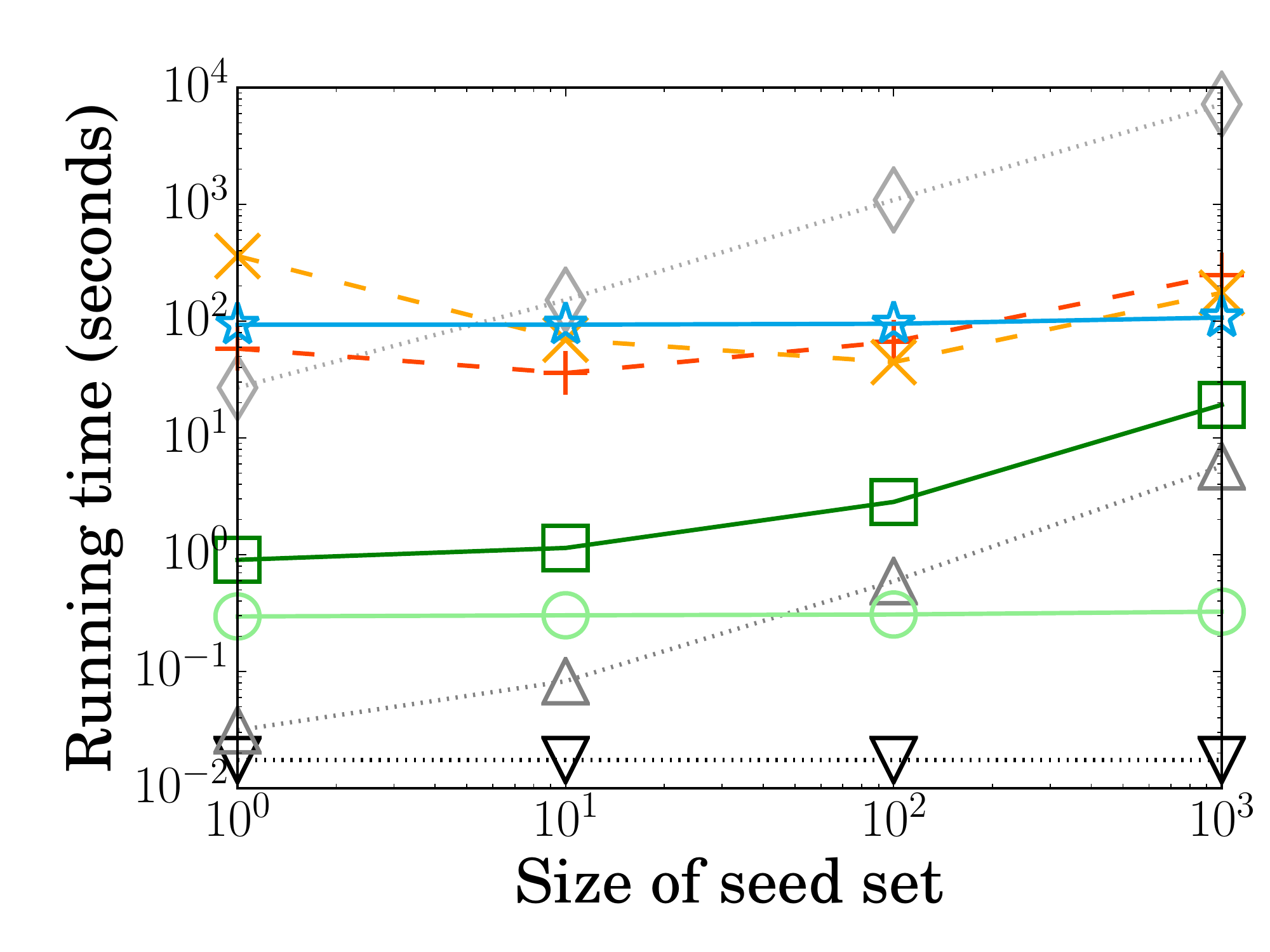}\label{subfig:livejournal_WC_time}}\hfill
	\subfloat[Twitter (IRIE cannot run)]{\includegraphics[width=0.31\linewidth]{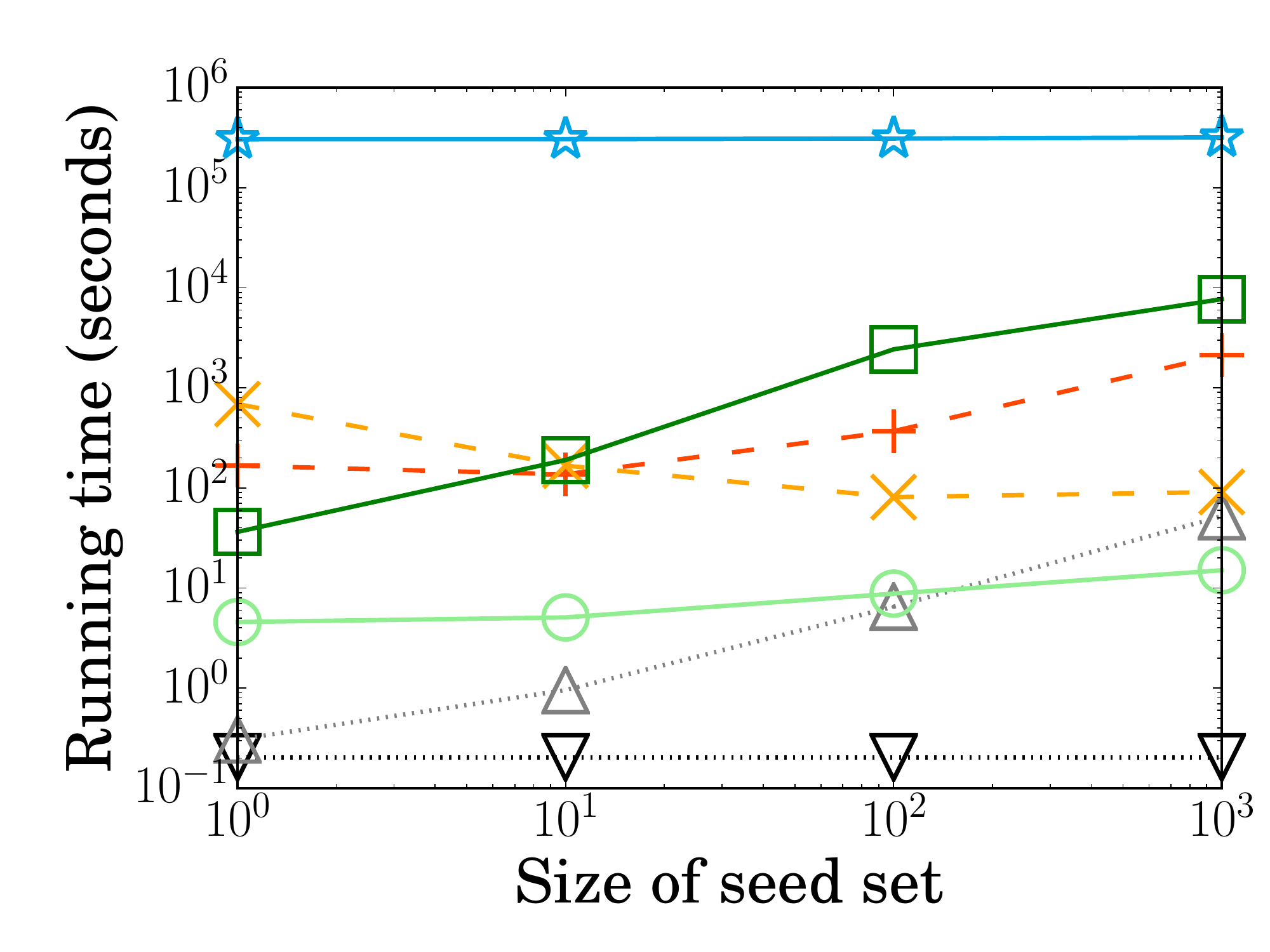}\label{subfig:big_twitter_WC_time}}
	\caption{Running time on various graphs under the WC model.}\label{fig:time_WC}
\end{figure*}

\begin{figure*}[!t]
	\capstart
	\centering
	\includegraphics[width=1.0\linewidth]{legend}\vspace{-0.15in}\\
	\subfloat[NetHEPT]{\includegraphics[width=0.31\linewidth]{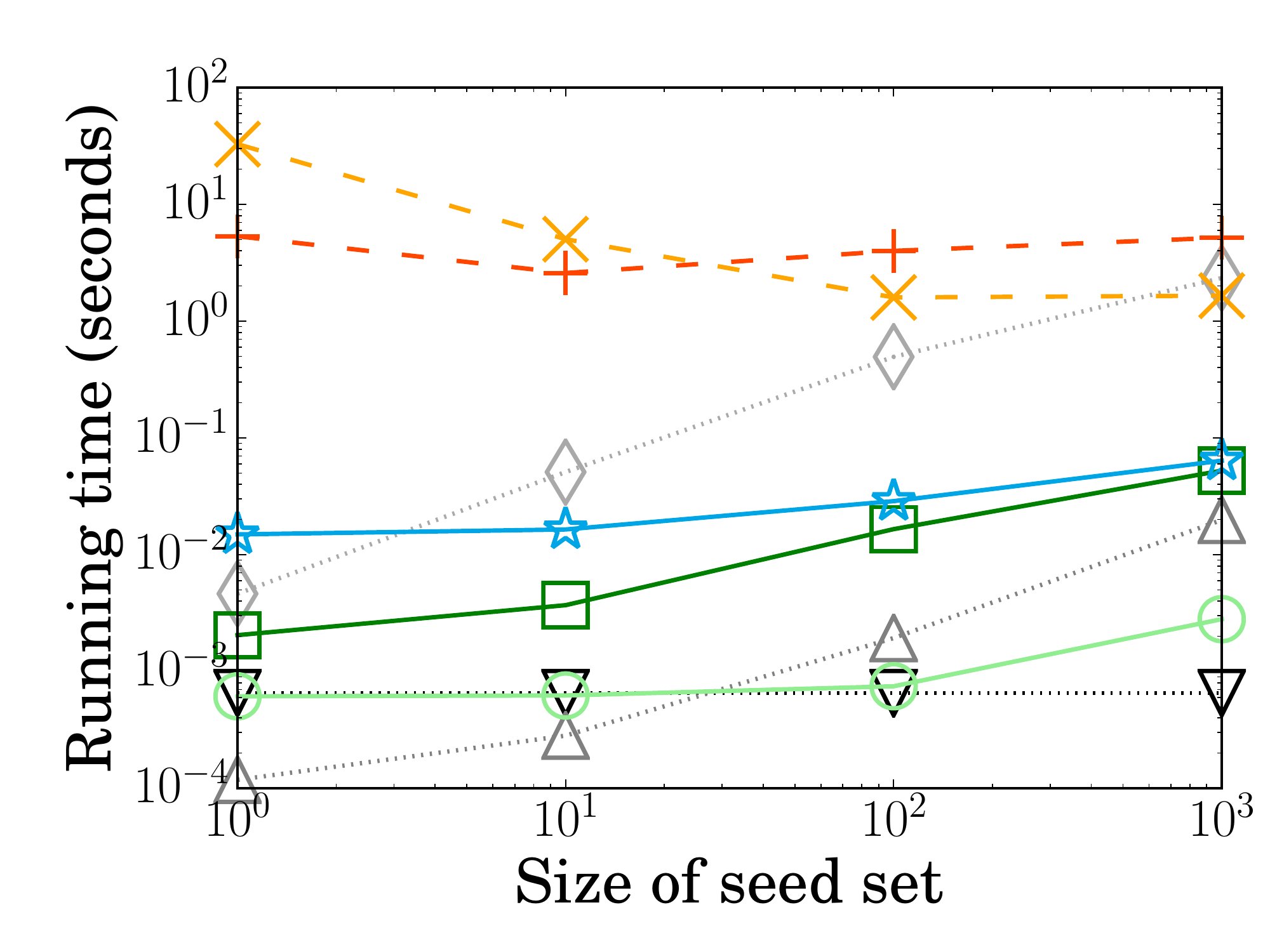}\label{subfig:NetHEPT_TR_time}}\hfill
	\subfloat[LiveJournal (IMM and D-SSA cannot run)
	]{\includegraphics[width=0.31\linewidth]{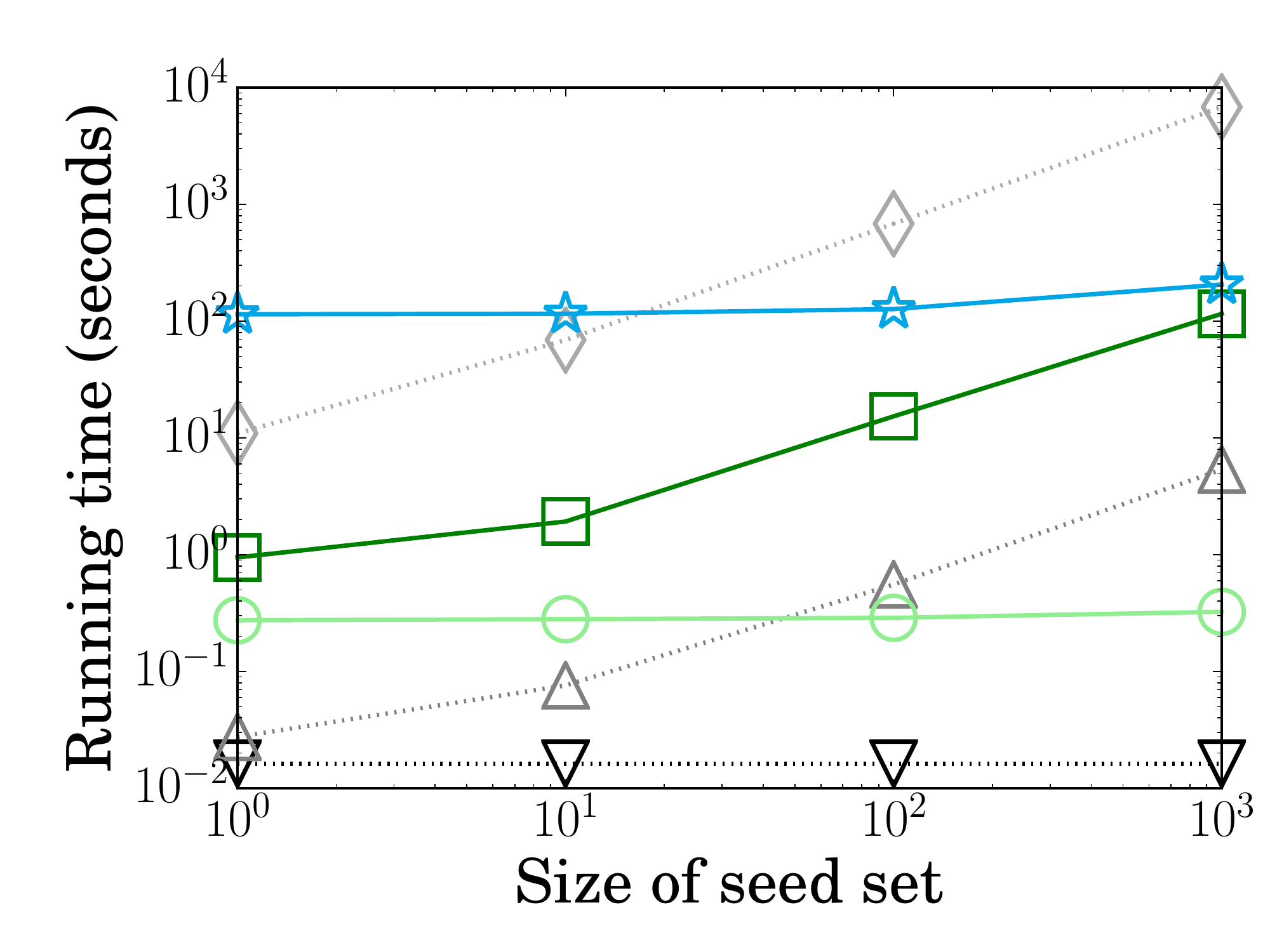}\label{subfig:livejournal_TR_time}}\hfill
	\subfloat[Twitter (IRIE, IMM and D-SSA cannot run)]{\includegraphics[width=0.31\linewidth]{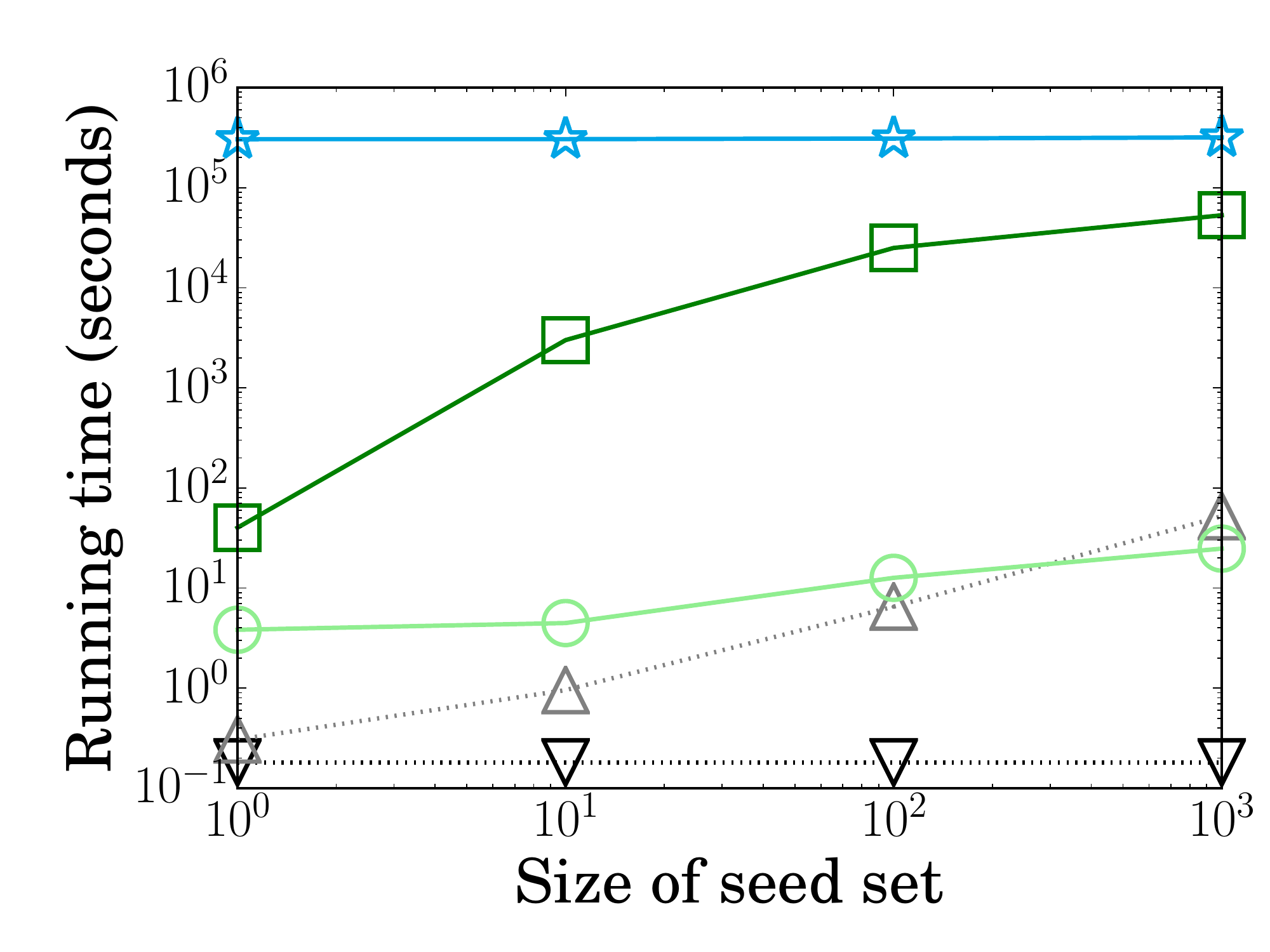}\label{subfig:big_twitter_TR_time}}
	\caption{Running time on various graphs under the TRIVALENCY model.}\label{fig:time_TR}
	\vspace{-0.1in}
\end{figure*}

\textbf{Running Time:} \figuresname~\ref{fig:time_WC} and \ref{fig:time_TR} show the running times of different algorithms. The OneHop and DegreeDiscount methods run almost at the same speed which can find the top $1000$ influential nodes on the Twitter dataset (with billions of edges) within $30$ seconds. They are just slightly slower than the HighDegree method and run several orders faster than other methods, including the IRIE, IMM, D-SSA, TwoHop and TwoHop-O methods (note that the y-axis is in logscale). This shows a tradeoff between efficiency and effectiveness for the hop-based methods. Estimating the influence spread with a higher hop limit takes more time but can improve the quality of the seed set chosen. For example, on the LiveJournal dataset under the WC model (\figurename~\ref{subfig:livejournal_WC_inf}), the TwoHop method performs notably better than the OneHop method. If the application is highly time-sensitive, the OneHop method could be preferable to the TwoHop method. Otherwise, the TwoHop method is favoured since its running time is quite acceptable even for very large networks. 

We also observe that while the TwoHop and TwoHop-O methods always produce the same seed set solution, the former runs significantly faster than the latter (by up to $4$ orders of magnitude). This is because TwoHop-O consumes too much time on computing the influence spread of all the single seed sets. This demonstrates the efficiency of our upper bounding approach. 

\begin{table}[!t]
	\centering
	\begin{threeparttable}
		\capstart
		\caption{Running time (seconds) for selecting 1000 seeds in the Twitter graph under the WC model with different propagation probability scale factors $f$.}
		\label{table:twitter}
		\setlength{\tabcolsep}{0.6em} 
		\renewcommand{\arraystretch}{1.1}
		\begin{tabular}{c||cccccc}
			\Xhline{1.2pt}
			Method  & $f=1.0$ & $1.1$ & $1.2$ & $1.3$ & $1.4$ & $1.5$ \\ \Xhline{0.5pt}
			HighDegree & 0.20 & 0.20 & 0.21 & 0.21 & 0.20 & 0.21 \\
			DegreeDiscount & 52.23 & 51.79 & 51.71 & 52.14 & 52.70 & 51.53 \\ 
			IRIE & \textendash & \textendash & \textendash & \textendash & \textendash & \textendash \\ \Xcline{4-7}{1.2pt}
			IMM & 2.13e3 & 3.52e4 & \multicolumn{1}{!{\vrule width1.2pt}c}{\textendash} & \textendash & \textendash & \multicolumn{1}{c!{\vrule width1.2pt}}{\textendash} \\
			D-SSA & 91.21 & 3.76e3 & \multicolumn{1}{!{\vrule width1.2pt}c}{2.44e4} & \textendash & \textendash & \multicolumn{1}{c!{\vrule width1.2pt}}{\textendash} \\
			OneHop & 15.07 & 16.08 & \multicolumn{1}{!{\vrule width1.2pt}c}{17.70} & 17.18 & 17.33 & \multicolumn{1}{c!{\vrule width1.2pt}}{17.85} \\
			TwoHop & 7.70e3 & 7.89e3 & \multicolumn{1}{!{\vrule width1.2pt}c}{8.11e3} & 8.25e3 & 7.94e3 & \multicolumn{1}{c!{\vrule width1.2pt}}{8.32e3}\\ \Xcline{4-7}{1.2pt}
			\Xhline{1.2pt}
		\end{tabular}
		\begin{tablenotes}
			\small
			\item The field with ``\textendash'' means that the method cannot run under the setting.
		\end{tablenotes}
	\end{threeparttable}
	\vspace{-0.1in}
\end{table}

Moreover, we observe that the OneHop and TwoHop methods generally run much faster than the state-of-the-art IRIE, IMM and D-SSA methods. This demonstrates the efficiency of our hop-based methods. Since IMM and D-SSA are reverse influence sampling methods, their running times heavily depend on the sizes of the reverse reachable sets sampled. Under the TRIVALENCY model, all the edges are likely to have substantial propagation probabilities. As a result, high-degree nodes can have many direct reverse reachable neighbors in a sample outcome of influence propagation. Thus, the samples of reverse reachable sets are large and require significant memory and time to compute, which makes IMM and D-SSA intractable on the LiveJournal and Twitter datasets. Under the WC model, for any node, the expected number of its direct reverse reachable neighbors is $1$ because the edge propagation probabilities are given by the reciprocal of the node's in-degree. This significantly limits the sizes of reverse reachable sets and favors IMM and D-SSA. However, the running times of IMM and D-SSA are still very sensitive to the propagation probabilities. Table~\ref{table:twitter} shows the trends when we scale the propagation probabilities under the WC model by a small factor up to $1.5$ (for selecting $1000$ seeds in the Twitter graph). It can be seen that the running times of IMM and D-SSA increase rapidly with the scale factor and far exceed our hop-based methods even at a minor factor of $1.2$. In contrast, our hop-based methods are much less sensitive to the propagation probabilities.

\textbf{Remark on Memory Usage:} Recall that our hop-based algorithms just need $O(|\mathcal{V}|)$ space which is negligible compared to the space $O(|\mathcal{V}|+|\mathcal{E}|)$ required for storing the OSN graph. On the other hand, the IRIE, IMM and D-SSA methods have significantly higher space complexities, e.g., those of IMM and D-SSA are both $O\Big(\frac{\big(\ln(1/\delta)+\ln\binom{|\mathcal{V}|}{k}\big)\cdot(|\mathcal{V}|+|\mathcal{E}|)}{\epsilon^{2}}\Big)$ \cite{Nguyen_DSSA_2016,Tang_IMM_2015}. Thus, they fail to produce results on very large datasets. Our hop-based algorithms never face the out-of-memory problems as long as the memory is large enough to store the OSN graph. 

\section{Related Work}\label{sec:relatedWork}

Since the $(1-1/e-\epsilon)$-approximation greedy algorithm was proposed by \cite{Kempe_maxInfluence_2003} for influence maximization, there has been considerable research on improving the efficiency of the greedy algorithm by using heuristics to trade the accuracy of influence estimation for computational efficiency \cite{Chen_MIA_2010,Chen_degreeDiscount_2009,Chen_LDAG_2010,Jung_IRIE_2012}, or optimizing the Monte-Carlo simulations for influence estimation \cite{Borgs_RIS_2014,Nguyen_BCT_2016,Nguyen_DSSA_2016,Ohsaka_prunedMC_2014,Tang_IMM_2015,Tang_reverse_2014}. Among them, the DegreeDiscount heuristic \cite{Chen_degreeDiscount_2009} roughly estimated influence spread within one-hop neighborhood and the PMIA heuristic \cite{Chen_MIA_2010} used independent propagation paths to construct arborescences for rough influence estimation. However, PMIA is very costly in both time and space compared to a follow-up IRIE algorithm \cite{Jung_IRIE_2012}. DegreeDiscount and IRIE as well as advanced sampling-based methods are all included in our experimental comparison. Leveraging the independency among propagation paths, we have developed efficient hop-based methods to compute the \textit{exact} influence spread within a certain number of hops.

Our hop-based influence estimation is in spirit similar to the time-constrained independent cascade model studied in \cite{Chen_time_2012,Dinh_CFM_2014,Liu_time_2012} by concentrating on the diffusion within a fixed number of hops. We make new technological advances by inventing algorithms to compute the exact influence spreads of one-hop and two-hop propagations, which could only be approximately estimated in previous work where the approximation guarantee is difficult to analyze \cite{Lee_hop_2014}. Our algorithms enable very efficient evaluation of the change in influence spread when a new seed node is added. We also derive an upper bound on the influence spread to further speed up our hop-based algorithms. Our hop-based approaches can be easily applied to many influence-based applications, such as topic-aware influence maximization \cite{Zhang_topic_2013} and community detection via influence maximization \cite{Jiang_community_2014}.

\section{Conclusion}\label{sec:conclusion}
In this paper, we have proposed lightweight hop-based methods to address the problem of influence maximization in OSNs. We have also developed an upper bounding technique to further speed up the seed selection algorithm. Through analysis, we show that our methods can provide certain theoretical guarantees. Experiments are conducted with real OSN datasets to compare the efficiency and effectiveness of our algorithms with state-of-the-art ones. In terms of solution quality, our hop-based methods are on par with the most advanced IMM and D-SSA methods which can provide the best $(1-1/e-\epsilon)$-approximation guarantee and remarkably outperform the HighDegree and DegreeDiscount heuristics for quite some cases. In terms of efficiency, our hop-based methods run much faster than the IRIE, IMM and D-SSA methods for most cases tested. Furthermore, while all these existing methods fail to run on some test cases, our hop-based methods can always execute and find solutions.

\section*{Acknowledgment}
This research is supported by the National Research Foundation, Prime Minister's Office, Singapore under its IDM Futures Funding Initiative, and by Singapore Ministry of Education Academic Research Fund Tier 1 under Grant 2013-T1-002-123.

\bibliographystyle{IEEEtranS}
\bibliography{reference}

\newpage
\appendix

\begin{IEEEproof}[Proof of Theorem~\ref{theorem:twohop_incremental}]
	To consider the outgoing edges from $u$ one at a time, we first disable all the edges from $u$ to its neighbors except for one edge $(u,w_1)$. Then, for each neighbor $v$ of $w_1$, all of $v$'s inverse neighbors other than $w_1$ have their one-hop activation probabilities unchanged by adding $(u,w_1)$. Let $\pi_2^{\mathcal{S}\cup\{u\}}(v|w_1)$ denote the new two-hop activation probability of $v$. Then, we have
	\begin{equation}
	\frac{1-\pi_2^{\mathcal{S}\cup\{u\}}(v|w_1)}{1-\pi_2^{\mathcal{S}}(v)}=\frac{1-p_{w_1,v}\cdot \pi_1^{\mathcal{S}\cup\{u\}}(w_1)}{1-p_{w_1,v}\cdot \pi_1^{\mathcal{S}}(w_1)}.
	\end{equation}
	Next, we enable the second edge $(u,w_2)$. Let $\pi_2^{\mathcal{S}\cup\{u\}}(v|w_1,w_2)$ denote the new two-hop activation probability of $v$. Following similar arguments, for each neighbor $v$ of $w_2$, we have
	\begin{equation}
	\frac{1-\pi_2^{\mathcal{S}\cup\{u\}}(v|w_1,w_2)}{1-\pi_2^{\mathcal{S}\cup\{u\}}(v|w_1)} =\frac{1-p_{w_2,v}\cdot \pi_1^{\mathcal{S}\cup\{u\}}(w_2)}{1-p_{w_2,v}\cdot \pi_1^{\mathcal{S}}(w_2)}.
	\end{equation}
	We continue to enable the outgoing edges of $u$ sequentially. In general, when an edge $(u,w_i)$ is enabled after edges $(u,w_1)$, $(u,w_2)$, $\cdots$, $(u,w_{i-1})$, for each neighbor $v$ of $w_i$, we have
	\begin{equation}
	\frac{1-\pi_2^{\mathcal{S}\cup\{u\}}(v|w_1,\cdots,w_i)}{1-\pi_2^{\mathcal{S}\cup\{u\}}(v|w_1,\cdots,w_{i-1})} =\frac{1-p_{w_i,v}\cdot \pi_1^{\mathcal{S}\cup\{u\}}(w_i)}{1-p_{w_i,v}\cdot \pi_1^{\mathcal{S}}(w_i)}.
	\end{equation}
	
	Therefore, we can initialize $\pi_2^{\mathcal{S}\cup\{u\}}(v)$ with $\pi_2^{\mathcal{S}}(v)$ and iteratively update $\pi_2^{\mathcal{S}\cup\{u\}}(v)$ with
	\begin{equation}\label{eq:update2}
	1-\big(1-\pi_2^{\mathcal{S}\cup\{u\}}(v)\big)\cdot\frac{1-p_{w,v}\cdot \pi_1^{\mathcal{S}\cup\{u\}}(w)}{1-p_{w,v}\cdot \pi_1^{\mathcal{S}}(w)},
	\end{equation}
	for all the nodes $w\in\mathcal{N}_u\setminus\mathcal{S}$ and $v\in\mathcal{N}_w\setminus\mathcal{S}$. Moreover, for the direct neighbors of $u$, their two-hop activation probabilities also need to be adjusted because $u$'s one-hop activation probability has changed from $\pi_1^{\mathcal{S}}(u)$ to $1$. For each neighbor $v$ of $u$, the adjustment can be made in a similar way by updating $\pi_2^{\mathcal{S}\cup\{u\}}(v)$ with
	\begin{equation}\label{eq:update1}
	1-\big(1-\pi_2^{\mathcal{S}\cup\{u\}}(v)\big)\cdot\frac{1-p_{u,v}\cdot \pi_1^{\mathcal{S}\cup\{u\}}(u)}{1-p_{u,v}\cdot \pi_1^{\mathcal{S}}(u)}.
	\end{equation}
	Then, the final two-hop activation probability $\pi_2^{\mathcal{S}\cup\{u\}}(v)$ by the iterative updates \eqref{eq:update2} and \eqref{eq:update1} is
	\begin{equation}
	\pi_2^{\mathcal{S}\cup\{u\}}(v) = 1-\big(1-\pi_2^{\mathcal{S}}(v)\big)\cdot\!\!\!\!\!\!\!\!\prod_{w\in(\mathcal{M}_{u,v}\cup\{u\})}\!\!\!\!\!\!\!\!\frac{1-p_{w,v}\cdot \pi_1^{\mathcal{S}\cup\{u\}}(w)}{1-p_{w,v}\cdot \pi_1^{\mathcal{S}}(w)}.
	\end{equation}
	
	Hence, the theorem is proven.
\end{IEEEproof}

\begin{IEEEproof}[Proof of Theorem~\ref{theorem:ubound}]
	Consider a single seed $\{u\}$. Let $\mathcal{A}_u\subseteq\mathcal{N}_u$ denote a subset of a node $u$'s neighbors. Let $p(\mathcal{A}_u)$ denote the probability that all the nodes in $\mathcal{A}_u$ are activated directly by $u$ in the IC model while all the nodes in $\mathcal{N}_u\setminus\mathcal{A}_u$ are not directly activated by $u$ (they may not even be activated eventually). Since each of $u$'s neighbors is activated by $u$ independently, we have
	\begin{equation}
	p(\mathcal{A}_u)=\big(\prod_{v\in\mathcal{A}_u} p_{u,v}\big)\cdot\big(\prod_{v\in\mathcal{N}_u\setminus\mathcal{A}_u}(1-p_{u,v})\big).
	\end{equation}
	Furthermore, with $h$ hops of propagation, for each node $w\in\mathcal{V}\setminus\{u\}$, $w$ can only be activated by a propagation path starting from a node $v\in\mathcal{A}_u$ whose path length is no longer than $h-1$ hops. In other words, the probability for $w$ to be activated by $\mathcal{A}_u$ is $\pi_{h-1}^{\mathcal{A}_u}(w)$. Considering all the possible node sets $\mathcal{A}_u$ activated directly by $u$, we have
	\allowdisplaybreaks[4]
	\begin{align*}
	&\sigma_h(\{u\})\\
	=\ & 1+\sum_{\mathcal{A}_u\subseteq\mathcal{N}_u}\Big(p(\mathcal{A}_u)\cdot\sum_{w\in\mathcal{V}\setminus\{u\}}\pi_{h-1}^{\mathcal{A}_u}(w)\Big)\\
	\leq\ & 1+\sum_{\mathcal{A}_u\subseteq\mathcal{N}_u}\Big(p(\mathcal{A}_u)\cdot\sum_{w\in\mathcal{V}}\pi_{h-1}^{\mathcal{A}_u}(w)\Big)\\
	=\ & 1+\sum_{\mathcal{A}_u\subseteq\mathcal{N}_u}\Big(p(\mathcal{A}_u)\cdot\sigma_{h-1}(\mathcal{A}_u)\Big)\\
	\leq\ & 1+\sum_{\mathcal{A}_u\subseteq\mathcal{N}_u}\Big(p(\mathcal{A}_u)\cdot\sum_{v\in\mathcal{A}_u}\sigma_{h-1}(\{v\})\Big)\\
	=\ & 1+\sum_{\mathcal{A}_u\subseteq\mathcal{N}_u}\Big(p(\mathcal{A}_u)\cdot\sum_{v\in\mathcal{N}_u}\big(\sigma_{h-1}(\{v\})\cdot p(v\in\mathcal{A}_u)\big)\Big)\\
	=\ & 1+\sum_{\mathcal{A}_u\subseteq\mathcal{N}_u}\Big(\sum_{v\in\mathcal{N}_u}\big(p(\mathcal{A}_u)\cdot\sigma_{h-1}(\{v\})\cdot p(v\in\mathcal{A}_u)\big)\Big)\\
	=\ & 1+\sum_{v\in\mathcal{N}_u}\Big(\sum_{\mathcal{A}_u\subseteq\mathcal{N}_u}\big(p(\mathcal{A}_u)\cdot\sigma_{h-1}(\{v\})\cdot p(v\in\mathcal{A}_u)\big)\Big)\\
	=\ & 1+\sum_{v\in\mathcal{N}_u}\Big(\sigma_{h-1}(\{v\})\cdot\sum_{\mathcal{A}_u\subseteq\mathcal{N}_u}\big(p(\mathcal{A}_u)\cdot p(v\in\mathcal{A}_u)\big)\Big).\numberthis\label{eq:sigma1}
	\end{align*}
	The second ``$\leq$'' is due to the submodularity of $\sigma_{h}(\cdot)$ (see Theorem~\ref{theorem:hop_influence}) such that $\sigma_{h-1}(\mathcal{A}_u)\leq \sum_{v\in\mathcal{A}_u}\sigma_{h-1}(\{v\})$. In the third ``='', $p(v\in\mathcal{A}_u)$ is such a binary value that $p(v\in\mathcal{A}_u)=1$ if and only if $v\in\mathcal{A}_u$. Meanwhile, we have
	\allowdisplaybreaks[4]
	\begin{align*}
	&\phantom{{}=}\sum_{\mathcal{A}_u\subseteq\mathcal{N}_u}\big(p(\mathcal{A}_u)\cdot p(v\in\mathcal{A}_u)\big)\\
	&=\sum_{\mathcal{A}_u\subseteq\mathcal{N}_u\setminus\{v\}}\big(p(\mathcal{A}_u)\cdot p(v\in\mathcal{A}_u)\big)\\
	&\phantom{{}=}+\sum_{\mathcal{A}_u\subseteq\mathcal{N}_u\setminus\{v\}}\big(p(\mathcal{A}_u\cup\{v\})\cdot p(v\in\mathcal{A}_u\cup\{v\})\big)\\
	&=\sum_{\mathcal{A}_u\subseteq\mathcal{N}_u\setminus\{v\}}p(\mathcal{A}_u\cup\{v\}).\numberthis\label{eq:probability1}
	\end{align*}
	The last ``='' follows the fact that $p(v\in\mathcal{A}_u)=0$ since $v\not\in\mathcal{A}_u\subseteq\mathcal{N}_u\setminus\{v\}$ and $p(v\in\mathcal{A}_u\cup\{v\})=1$ since $v\in\mathcal{A}_u\cup\{v\}$. Therefore, from \eqref{eq:sigma1} and \eqref{eq:probability1}, we have
	\begin{equation}\label{eq:probability2}
	\sigma_h(\{u\})\leq 1+\sum_{v\in\mathcal{N}_u}\Big(\sigma_{h-1}(\{v\})\cdot\sum_{\mathcal{A}_u\subseteq\mathcal{N}_u\setminus\{v\}}p(\mathcal{A}_u\cup\{v\})\Big).
	\end{equation}
	Furthermore, by definition,
	\allowdisplaybreaks[4]
	\begin{align*}
	&\!\!\!\!\sum_{\mathcal{A}_u\subseteq\mathcal{N}_u\setminus\{v\}}\!\!\!\!p(\mathcal{A}_u\cup\{v\})\\
	=\ &\!\!\!\!\sum_{\mathcal{A}_u\subseteq\mathcal{N}_u\setminus\{v\}}\!\!\!\!\Big(\big(\!\!\prod_{w\in\mathcal{A}_u\cup\{v\}}\!\! p_{u,w}\big)\cdot\big(\!\!\!\!\!\!\prod_{w\in\mathcal{N}_u\setminus(\mathcal{A}_u\cup\{v\})}\!\!\!\!\!\!(1-p_{u,w})\big)\Big)\\
	=\ &\!\!\!\!\sum_{\mathcal{A}_u\subseteq\mathcal{N}_u\setminus\{v\}}\!\!\!\!\Big(p_{u,v}\cdot\big(\prod_{w\in\mathcal{A}_u} p_{u,w}\big)\cdot\big(\!\!\!\!\!\!\prod_{w\in\mathcal{N}_u\setminus(\mathcal{A}_u\cup\{v\})}\!\!\!\!\!\!(1-p_{u,w})\big)\Big)\\
	=\ & p_{u,v}\cdot\!\!\!\!\sum_{\mathcal{A}_u\subseteq\mathcal{N}_u\setminus\{v\}}\!\!\!\!\Big(\big(\prod_{w\in\mathcal{A}_u} p_{u,w}\big)\cdot\big(\!\!\!\!\!\!\prod_{w\in\mathcal{N}_u\setminus(\mathcal{A}_u\cup\{v\})}\!\!\!\!\!\!(1-p_{u,w})\big)\Big)\\
	=\ & p_{u,v}\cdot 1\\
	=\ & p_{u,v}.\numberthis\label{eq:probability3}
	\end{align*}
	Thus, by \eqref{eq:probability2} and \eqref{eq:probability3}, it holds that $\sigma_h(\{u\})\leq 1+\sum_{v\in\mathcal{N}_u}\Big(\sigma_{h-1}(\{v\})\cdot p_{u,v}\Big)$.
	
	Inequality (\ref{eq:ubound2}) can be proved by induction. When $h=1$, the inequality follows directly from Inequality (\ref{eq:ubound1}). Suppose that it holds for $h-1$ hops of propagation, i.e., $\sigma_{h-1}(\{u\}) \leq \hat{\sigma}_{h-1}(\{u\})$. Then, for $h$ hops of propagation, we have
	\begin{align*}
	\sigma_h(\{u\})
	&\leq 1+\sum_{v\in\mathcal{N}_u}\Big(p_{u,v}\cdot \sigma_{h-1}(\{v\})\Big)\\
	&\leq 1+\sum_{v\in\mathcal{N}_u}\Big(p_{u,v}\cdot \hat{\sigma}_{h-1}(\{v\})\Big)\\
	&= \hat{\sigma}_{h}(\{u\}).\numberthis
	\end{align*}
	Therefore, for any $h\geq 0$, we have $\sigma_{h}(\{u\}) \leq \hat{\sigma}_{h}(\{u\})$.
\end{IEEEproof}

\begin{IEEEproof}[Proof of Theorem~\ref{theorem:hop_influence}]
	This can be proved using a similar approach to that by Kempe \textit{et al.} in \cite{Kempe_maxInfluence_2003}. For each edge $(u,v)\in\mathcal{E}$, we independently flip a coin of bias $p_{u,v}$ to decide whether the edge $(u,v)$ is \textit{live} or \textit{blocked} to generate a sample influence propagation outcome $X$. We use $p(X)$ to denote the probability of a specific outcome $X$ in the sample space. Let $\mathcal{V}_h^X(v)$ denote the node set that can be reached from a node $v$ within $h$ hops in the sample outcome $X$. Then, the number of nodes that can be reached from a seed set $\mathcal{S}$ within $h$ hops in the outcome $X$ is given by $\sigma_h^X(\mathcal{S})=\Big{|}\bigcup_{v\in\mathcal{S}}\mathcal{V}_h^X(v)\Big{|}$.
	Thus,
	\begin{equation}
	\sigma_h(\mathcal{S})=\sum_{X}\big(p(X)\cdot\sigma_h^X(\mathcal{S})\big),
	\end{equation}
	where the monotonicity of $\sigma_h(\mathcal{S})$ holds since $\sigma_h^X(\mathcal{S})$ increases as $\mathcal{S}$ expands.
	
	The marginal influence gain 
	\begin{equation}
	\sigma_h^X(\mathcal{S}\cup\{u\})-\sigma_h^X(\mathcal{S})=\Big{|}\mathcal{V}_h^X(u)\setminus\bigcup_{v\in\mathcal{S}}\mathcal{V}_h^X(v)\Big{|}
	\end{equation}
	is the number of nodes that are reachable from a node $u$ within $h$ hops but are not reachable from any node in a seed set $\mathcal{S}$ within $h$ hops in a sample outcome $X$. For any two node sets $\mathcal{S}$ and $\mathcal{T}$ where $\mathcal{S}\subseteq\mathcal{T}$, we have $\bigcup_{v\in\mathcal{S}}\mathcal{V}_h^X(v)\subseteq\bigcup_{v\in\mathcal{T}}\mathcal{V}_h^X(v)$.
	Thus, $\mathcal{V}_h^X(u)\setminus\bigcup_{v\in\mathcal{S}}\mathcal{V}_h^X(v)\supseteq \mathcal{V}_h^X(u)\setminus\bigcup_{v\in\mathcal{T}}\mathcal{V}_h^X(v)$,
	which implies that 
	\begin{equation}
	\sigma_h^X(\mathcal{S}\cup\{u\})-\sigma_h^X(\mathcal{S})\geq \sigma_h^X(\mathcal{T}\cup\{u\})-\sigma_h^X(\mathcal{T}).
	\end{equation}
	Since $p(X)\geq 0$ for any $X$, taking the linear combination, we have 
	\begin{equation}
	\sigma_h(\mathcal{S}\cup\{u\})-\sigma_h(\mathcal{S})\geq \sigma_h(\mathcal{T}\cup\{u\})-\sigma_h(\mathcal{T}).
	\end{equation}
	Thus, $\sigma_h(\cdot)$ is submodular.
\end{IEEEproof}

\begin{IEEEproof}[Proof of Theorem~\ref{theorem:ratio_guarantee}]
	Let $\mathcal{S}_h^*$ denote the optimal seed set for maximizing the influence within $h$ hops of propagation, i.e., $\sigma_h(\mathcal{S}_h^*)=\max_{|\mathcal{S}|=k}\sigma_h(\mathcal{S})$. We have $\sigma(\mathcal{S}_h)\geq\sigma_h(\mathcal{S}_h)\geq(1-1/e)\sigma_h(\mathcal{S}_h^*)\geq(1-1/e)\sigma_h(\mathcal{S}^*)\geq\big((1-1/e)\alpha\big)\cdot\sigma(\mathcal{S}^*)$. The first inequality follows from the fact that the exact influence spread is equal to the influence spread without any hop limitation of propagation. The second inequality is due to the submodularity and monotonicity of $\sigma_h(\cdot)$ (Theorem~\ref{theorem:hop_influence}). The third inequality is because $\mathcal{S}_h^*$ is the optimal solution for maximizing $\sigma_h(\cdot)$. The last inequality is by the assumption that $\sigma_h(\mathcal{S})/\sigma(\mathcal{S})\geq\alpha$ for any $|\mathcal{S}|=k$.
\end{IEEEproof}

We first introduce some lemmas used to prove Theorem~\ref{theorem:ratio}.
\begin{lemma}\label{lemma:one_hop_influence}
	For scale free random graphs with propagation probability $p_{u,v}=p$ for every edge $(u,v)\in\mathcal{E}$, the expected influence spread produced within one hop of propagation from a random seed set $\mathcal{S}$ satisfies
	\begin{equation}\label{eq:one_hop_influence}
	\mathbb{E}[\sigma_1(\mathcal{S})]\geq(p+1)k-pk^2/|\mathcal{V}|.
	\end{equation}
\end{lemma}

\begin{IEEEproof}[Proof of Lemma~\ref{lemma:one_hop_influence}]
	With one hop of propagation, for a randomly selected node $v$, it is not activated if and only if $v$ is not a seed and $v$ is not activated by any of its inverse neighbors. The probability for $v$ to be a non-seed node is $1-\frac{k}{|\mathcal{V}|}$. The probability for an inverse neighbor of $v$ to be a seed is $\frac{k}{|V|}$ and thus, the probability for it to activate $v$ is $p\cdot\frac{k}{|\mathcal{V}|}$. Therefore, the probability for all of $v$'s inverse neighbors to fail to activate $v$ is
	\begin{equation}
	\prod_{u\in\mathcal{I}_v}\big(1-p\cdot\frac{k}{|\mathcal{V}|}\big)=\big(1-\frac{pk}{|\mathcal{V}|}\big)^{|\mathcal{I}_v|}.
	\end{equation}
	Note that if $v$ is selected as a seed, it must be activated. Hence, the overall activation probability of $v$ is
	\begin{equation}
	\pi_1^\mathcal{S}(v)=1-\big(1-\frac{k}{|\mathcal{V}|}\big)\cdot\big(1-\frac{pk}{|\mathcal{V}|}\big)^{|\mathcal{I}_v|}.
	\end{equation}
	As a result, the expectation of the activation probability of a random node $v$ is given by
	\allowdisplaybreaks[4]
	\begin{align*}
	\mathbb{E}[\pi_1^\mathcal{S}(v)]
	&=\mathbb{E}\Big[1-\big(1-\frac{k}{|\mathcal{V}|}\big)\cdot\big(1-\frac{pk}{|\mathcal{V}|}\big)^{|\mathcal{I}_v|}\Big]\\
	&=1-\big(1-\frac{k}{|\mathcal{V}|}\big)\cdot\sum_{|\mathcal{I}_v|=1}^{\infty}\Big(P_0(|\mathcal{I}_v|)\cdot\big(1-\frac{pk}{|\mathcal{V}|}\big)^{|\mathcal{I}_v|}\Big)\\
	&\geq 1-\big(1-\frac{k}{|\mathcal{V}|}\big)\cdot\big(1-\frac{pk}{|\mathcal{V}|}\big)\cdot\sum_{|\mathcal{I}_v|=1}^{\infty}P_0(|\mathcal{I}_v|)\\
	&=1-\big(1-\frac{k}{|\mathcal{V}|}\big)\cdot\big(1-\frac{pk}{|\mathcal{V}|}\big)\\
	&=\frac{(1+p)k}{|\mathcal{V}|}-\frac{pk^2}{|\mathcal{V}|^2}.\numberthis
	\end{align*}
	
	Therefore, it holds that $\mathbb{E}[\sigma_1(\mathcal{S})]=|\mathcal{V}|\cdot\mathbb{E}[\pi_1^\mathcal{S}(v)]\geq(p+1)k-pk^2/|\mathcal{V}|$. This completes the proof.
\end{IEEEproof}

\begin{lemma}[\cite{Li_advertising_2012}]\label{lemma:expecated_fraction}
	For an infinite random power law graph, the expected fraction of nodes activated $\phi(\mathcal{S})=\mathbb{E}[\sigma(\mathcal{S})]/|\mathcal{V}|$ can be computed by 
	\begin{equation}\label{eq:expecated_fraction}
	\begin{cases}
	1-\varphi(\mathcal{S})=\big(1-\frac{k}{|\mathcal{V}|}\big)\sum_{d=0}^{\infty}P_1(d+1)\big(1-p\varphi(\mathcal{S})\big)^d,\vspace{0.05in}\\
	1-\phi(\mathcal{S})=\big(1-\frac{k}{|\mathcal{V}|}\big)\sum_{d=1}^{\infty}P_0(d)\big(1-p\varphi(\mathcal{S})\big)^d,
	\end{cases}
	\end{equation}
	where $P_1(d)=\frac{d^{1-\gamma}}{\sum_{d_i=1}^{\infty}d_i^{1-\gamma}}$ is the probability of a node connecting to a neighbor whose degree is $d$, and $\varphi(\mathcal{S})$ is an instrumental variable.
\end{lemma}

\begin{lemma}\label{lemma:expecated_fraction_bound}
	The expected fraction of nodes activated $\phi(\mathcal{S})$ is bounded by 
	\begin{equation}\label{eq:expected_influence_infinity}
	\mathbb{E}[\sigma(\mathcal{S})]\leq |\mathcal{V}|\cdot\Big(1-\big(1-\frac{k}{|\mathcal{V}|}\big)P_0(1)(1-pA)\Big),
	\end{equation}
	where $A=1-\big(1-\frac{k}{|\mathcal{V}|}\big)P_1(1)$.
\end{lemma}

\begin{IEEEproof}[Proof of Lemma~\ref{lemma:expecated_fraction_bound}]
	From \eqref{eq:expecated_fraction} in Lemma~\ref{lemma:expecated_fraction}, we have
	\begin{equation}\label{eq:expecated_fraction_bound1}
	1-\varphi(\mathcal{S})\geq\big(1-\frac{k}{|\mathcal{V}|}\big)P_1(1)\big(1-p\varphi(\mathcal{S})\big)^0=1-A,
	\end{equation}
	and
	\begin{equation}\label{eq:expecated_fraction_bound2}
	1-\phi(\mathcal{S})\geq\big(1-\frac{k}{|\mathcal{V}|}\big)P_0(1)\big(1-p\varphi(\mathcal{S})\big).
	\end{equation}
	Hence, by \eqref{eq:expecated_fraction_bound1} and \eqref{eq:expecated_fraction_bound2}, the lemma follows.
\end{IEEEproof}

\begin{IEEEproof}[Proof of Theorem~\ref{theorem:ratio}]
	Lemma~\ref{lemma:one_hop_influence} indicates that 
	\begin{equation}\label{eq:lower}
	\mathbb{E}[\sigma_h(\mathcal{S})]\geq\mathbb{E}[\sigma_1(\mathcal{S})]\geq(p+1)k-pk^2/|\mathcal{V}|.
	\end{equation}
	Lemma~\ref{lemma:expecated_fraction_bound} indicates that
	\begin{equation}\label{eq:upper}
	E[\sigma(\mathcal{S})]\leq |\mathcal{V}|\cdot\Big(1-\big(1-\frac{k}{|\mathcal{V}|}\big)P_0(1)(1-pA)\Big).
	\end{equation}
	Putting \eqref{eq:lower} and \eqref{eq:upper} together, the theorem follows. 
\end{IEEEproof}
\balance
\end{document}